\documentclass[12pt, a4paper, reqno]{amsart}
\usepackage{mathrsfs}
\usepackage{amsmath}
\usepackage{amssymb}
\usepackage[all]{xy}

\newcommand{\liea}{\mathfrak {g}}
\newcommand{\liek}{\mathfrak {k}}
\newcommand{\lieG}{\mathcal{L}ie(\mathscr{G})}
\newcommand{\Ad}{\text{Ad}(P)}
\newcommand{\ad}{\text{ad}(P)}

\newcommand{\FF}{\mathbb {F}}
\newcommand{\da}{\mathscr{D}\mathscr{A}}
\newtheorem{prop}{Proposition}
\newtheorem{lema}[prop]{Lemma}

\newtheorem{defi}[prop]{Definition}
\newtheorem{rem}[prop]{Remark}
\newcommand{\mr}[1]{\mathscr{#1}}
\renewcommand{\AA}{\mathbb{A}}

\renewcommand{\AA}{\mathbb{A}}

\newcommand{\bs}[1]{\boldsymbol{#1}}
\newcommand{\re}{\mathbb{R}}

\title{Extended Connection in Yang-Mills Theory}
\author[G. Catren and J. Devoto]{\textbf{Gabriel Catren}\email{gabrielcatren@yahoo.com.ar} \\\\ \emph{Instituto de Astronom\'{i}a y F\'{i}sica del Espacio \\
C.C. 67, Sucursal 28, 1428, Buenos  Aires, Argentina} \\\\
and \\\\ \textbf{Jorge Devoto}\email{jdevoto@dm.uba.ar} \\\\
\emph{Math. Dept. FCEN, UBA, Ciudad Universitaria, pabellon 1,
1428, Buenos Aires, Argentina}}

\begin{document}

\maketitle
\begin{abstract}
The three fundamental geometric components of Yang-Mills theory
–-gauge field, gauge fixing and ghost field–- are unified in a new
object: an extended connection in a properly chosen principal
fiber bundle. To do this, it is necessary to generalize the notion
of gauge fixing by using a gauge fixing connection instead of a
section. From the equations for the extended connection's
curvature, we derive the relevant BRST transformations without
imposing the usual horizontality conditions. We show that the
gauge field's standard BRST transformation is only valid in a
local trivialization and we obtain the corresponding global
generalization. By using the Faddeev-Popov method, we apply the
generalized gauge fixing to the path integral quantization of
Yang-Mills theory. We show that the proposed gauge fixing can be
used even in the presence of a Gribov's obstruction.
\end{abstract}

\newpage

\section{Introduction}
\label{sec:0}

It is our objective in the present work to show how the main
geometric structures of Yang-Mills theory can be unified in a
single geometric object, namely a \emph{connection} in an infinite
dimensional principal fiber bundle. We will show how this
geometric formalism can be useful to the path integral
quantization of Yang-Mills theory. Some of the historical
motivations for the study of an extended connection in Yang-Mills
theory are the following. In the beginning of the 80's Yang Mills
theory was at the center of important mathematical developments,
especially Donaldson's theory of four manifolds'
invariants~\cite{donaldson90} and Witten's interpretation of this
theory in terms of a topological quantum field
theory~\cite{Witten:1988ze} (for a general review on topological
field theories see Refs.\cite{Blau91}, \cite{cordes}). A central
aspect of these theories is the study of the topological
properties of the space $\mr{A} / \mr{G}$, where $\mr{A}$ is the
configuration space of connections in a $G$-principal bundle $P
\rightarrow M$ and $\mr{G}$ the \emph{gauge group} of vertical
automorphisms of $P$. It is possible to show that under certain
hypotheses one obtains a $\mr{G}$-principal bundle structure
$\mr{A} \rightarrow \mr{A}/ \mr{G}$ \cite{donaldson90}. The
non-triviality of many invariants is then intimately linked with
the topological non-triviality of this bundle. In
Ref.\cite{Baulieu-Singer1988} Baulieu and Singer showed that
Witten's theory can be interpreted in terms of the gauge fixed
version of a topological action through a standard BRST procedure.
To do so, the authors unify the gauge field $A$ and the ghost
field $c$ for Yang-Mills symmetry in an extended connection
$\omega = A + c$ defined in a properly chosen principal bundle.
The curvature $\mr{F}$ of $\omega$ splits naturally as $\mr{F}=F +
\psi + \phi $, where $\psi$ and $\phi$ are the ghost for the
topological symmetry and the ghost for ghost respectively (the
necessity of a ghost for ghost is due to the dependence of the
topological symmetry on the Yang-Mills symmetry). By expanding the
expressions for the curvature $\mr{F}$ and the corresponding
Bianchi identity, the BRST transformations for this topological
gauge theory are elegantly recovered. It is commonly stated that
the passage from this topological Yang-Mills theory to the
ordinary (i.e. non-topological) case is mediated by the
horizontality (or flatness) conditions, i.e. by the conditions
$\psi= \phi =0$ (see Refs.\cite{Baulieu-Bellon},
\cite{Baulieu-Mieg1982}, \cite{Mieg}). In this case, an extended
connection $\omega = A + c$ can also be defined for an ordinary
Yang-Mills theory \emph{with a horizontal curvature of the form
$\mr{F}=F$}.

In this work, an extended connection $\AA$ for an ordinary
Yang-Mills theory will be defined as the sum of two factors. The
first factor is a universal family $\mathbf{A}^{U}$ parameterized
by $\mr{A}$ of connections in the $G$-principal bundle $P
\rightarrow M$. The second factor is an arbitrarily chosen
connection $\eta$ in the $\mr{G}$-principal fiber bundle $\mr{A}
\rightarrow \mr{A}/\mr{G}$. We will show that the connection
$\eta$ encodes the ghost field that generates the BRST complex. In
a local trivialization the ghost field can be identified with the
canonical vertical part of the connection $\eta$, which
corresponds to the Maurer-Cartan form of the gauge group $\mr{G}$
\cite{bonora83}. Alternatively, the ghost field can be considered
a universal connection in the gauge group's Weil algebra. The
connection $\eta$ can then be defined as the image of this
universal connection under a particular Chern-Weil homomorphism.
We will then argue that the connection $\eta$ also defines what
might be called \emph{generalized gauge fixing}. In fact, the
connection $\eta$ defines a horizontal subspace at each point of
the fiber bundle $\mr{A} \rightarrow \mr{A}/\mr{G}$. These
subspaces can be considered first order infinitesimal germs of
sections. The significant difference is that the connection $\eta$
induces a global section $\sigma_{\eta}$ in a fiber bundle
associated to the \emph{space of paths} in $\mr{A} /\mr{G}$. In
fact, the section $\sigma_{\eta}$ assigns the horizontal lift
defined by $\eta$ to each path $\left[\gamma\right]\subset \mr{A}
/\mr{G}$. Since the path integral is not an integral in the
\emph{space of fields} $\mr{A}$, but rather an integral in the
\emph{space of paths} in $\mr{A}$, the section $\sigma_{\eta}$
allows us to eliminate the gauge group's infinite volume in the
corresponding path integral. We will then show that the gauge
fixed action corresponding to the generalized gauge fixing defined
by $\eta$ can be obtained by means of the usual Faddeev-Popov
method. One of the main advantages over the usual formulation is
that the connection $\eta$ is globally well-defined even when the
topology of the fiber bundle $\mr{A} \rightarrow \mr{A}/\mr{G}$ is
not trivial (Gribov's obstruction). In this way, the existence of
a generalized notion of gauge fixing demonstrates that the
Faddeev-Popov method is still valid even in the presence of a
Gribov's obstruction. It is worth stressing that the extended
connection $\AA$ encodes not only the gauge and ghost fields (as
the connection $\omega = A + c$ in Ref.\cite{Baulieu-Singer1988}),
but also the definition of a global gauge fixing of the theory. We
will then show that the proposed formalism allows us to recover
the BRST transformations of the relevant fields without imposing
the usual horizontality conditions (\cite{Baulieu-Bellon},
\cite{Baulieu-Mieg1982}, \cite{Mieg}). This implies that the
extended connection's curvature $\FF$ does not necessarily have
the horizontal form $\FF=F$. Moreover, we will show that the gauge
field's standard BRST transformation is valid only in a local
trivialization of the fiber bundle $\mr{A} \rightarrow
\mr{A}/\mr{G}$. We will then find the corresponding global
generalization.

The paper is organized as follows. In section~\ref{sec:2} we
define the extended connection. In section~\ref{sec:3} we study
the curvature of this connection and show how the curvature forms
induce the BRST transformations of the different fields. In
section~\ref{sec:4} we study the relation between the gauge fixing
connection and the ghost field. In
section~\ref{sec:quant-hamilt-sett} we define the generalized
gauge fixing at the level of path integrals. In
section~\ref{sec:path-integr-quant} we calculate the gauge fixed
action by means of the Faddeev-Popov method. In the final section
we summarize the proposed formalism.

\section{Extended Connection}
\label{sec:2}

Let $\mathcal{M}$ denote space-time. We will suppose that it is
possible to define a foliation of $\mathcal{M}$ by spacelike
hypersurfaces. This foliation is defined by means of a
diffeomorphism $\iota: \mathcal{M} \rightarrow \mathbb{R} \times
M$, where $M$ is a smooth 3-dimensional Riemannian manifold. We
will also assume for the sake of simplicity that $M$ is compact.
Let $G$ be a compact Lie group with a fixed invariant inner
product in its Lie algebra $\liea$ and let $\mathcal{P}
\rightarrow \mathcal{M}$ be a fixed $G$-principal bundle. Using
the diffeomorphism $\iota$ and the fact that $\mathbb{R}$ is
contractible, we can assume that the fiber bundle $\mathcal{P}
\rightarrow \mathcal{M}$ over the space-time $\mathcal{M}$ is the
pullback of a fixed $G$-principal bundle $P \rightarrow M$ over
the space $M$
\begin{equation*}
  \begin{xy}
    \xymatrix{
      *{\mathcal{P} \cong p_{st}^* (P)} \ar[d] \ar[r] & P \ar[d] \\
      \mathcal{M} \ar[r]^{p_{st}} & M,
      }
  \end{xy}
\end{equation*}
where $p_{st}: \mathcal{M}\simeq \mathbb{R} \times M \rightarrow
M$ is the projection onto the second factor. We will denote by
$\Ad$ the fiber bundle $P \times_G G \rightarrow M$ associated to
the \emph{adjoint} action of $G$ on itself and by $\ad$ the vector
bundle $P \times_G \liea \rightarrow M$ associated to the adjoint
representation of $G$ on $\liea$. The \emph{gauge group} $\mr{G}$
is the group of vertical automorphisms of $P$. It can be naturally
identified with the space of sections of $\Ad $. Its Lie algebra
$\lieG$ is the space of sections of $\ad$. Its elements can be
identified with $G$-equivariant maps $\boldsymbol{\mathfrak{g}}: P
\rightarrow \liea$.

In the case of a principal fiber bundle over a finite dimensional
manifold with a compact structure group, there are three
equivalent definitions of connections~\cite[Chapter
2]{donaldson90}. In what follows, we will also consider
connections on infinite dimensional spaces (see
Refs.\cite{kriegl97}, \cite{Michor91}). For the general case, we
will use the following as the basic definition
~\cite{kobayashi63:_found}.
\begin{defi}
  Let $K$ be a Lie group with Lie algebra $\liek$ and let $E
  \xrightarrow{\pi} X$ be a $K$-principal bundle over a manifold $X$
  (both $K$ and $X$ can have infinite dimension). A \emph{connection}
  on $E$ is an \emph{equivariant distribution} $H$,
  i.e. a smooth field of vector spaces $H_p \subset TE_p$ (with $p \in
  E$) such that
 \begin{enumerate}
 \item For all $p \in E$ there is a direct sum decomposition
   \begin{equation}
     \label{eq:21}
     TE_p = H_p \oplus \text{ker} \, d\pi_p.
   \end{equation}
 \item The field is preserved by the induced action of $K$ on
 $TE$, i.e. $$H_{pg}=R_{g^{\ast}}H_{p},$$ where $R_{g^{\ast}}$ denotes
 the differential of the right translation by $g\in K$.
 \end{enumerate}
\end{defi}

As in the finite dimensional case, we can assign to each
connection a $K$-equivariant $\liek$-valued 1-form $\omega$ on $E$
such that $H_p=\text{ker } \omega$. The action of $K$ on $\liek$
is the adjoint action.

A connection can also be considered a $K$-invariant splitting of
the exact sequence
 \begin{equation}
   \label{eq:24}
   \begin{xy}
 \xymatrix{
   0 \ar[r]  & \text{ker} \, d\pi \ar[r]_{\iota} & TE  \ar[r]_{d\pi}
   & \pi^* TX \ar[r] \ar@/_1pc/[l]_{\sigma}  & 0,
   }
\end{xy}
 \end{equation}
where $\pi^* TX \rightarrow E$ is the pullback induced by the
projection $\pi: E \rightarrow X$ of the bundle $TX \rightarrow
X$. In this sequence $H_p = \sigma (\pi^{\ast}T_{\pi(p)} X)$.
Given a connection $\omega$ on $E$ the splitting considered in
equation~(\ref{eq:24}) induces  an isomorphism $ TE \simeq \pi^*
TX \oplus \text{ker} \, d\pi $.

The bundle $T_V E\doteq \text{ker} \, d\pi $ is called the bundle
of \emph{vertical tangent vectors}. This bundle is intrinsically
associated to the definition of principal bundles. Given a
connection, the isomorphism $TE \simeq \pi^* TX \oplus \text{ker}
\, d\pi $ induces a projection $\Pi_V^{\omega}: TE \rightarrow T_V
E$. By definition, the \emph{vertical cotangent bundle} $T_V^* E $
is the annihilator of $\sigma \left( \pi^* TX \right)$. In other
words, a $k$-form $\alpha$ is vertical if and only if it vanishes
whenever one of its arguments is a vector in $\sigma \left( \pi^*
TX \right)$. It is worth noticing that the definition of the
bundle $T_V^* E $ requires a connection. The sections of
$\Omega_V^k (E)\doteq \wedge^k T_V^* E $ are called
\emph{vertical} $k$-\emph{forms}. By definition, the connection
form $\omega$ is a vertical form.

Given a connection there is a decomposition of the de Rham
differential on $E$
\begin{equation}
  \label{eq:29}
  d_E = d_H + d_V
\end{equation}
into a horizontal and a vertical part. The horizontal part
corresponds to the \emph{covariant derivative}. The vertical part
is defined by the expression
\begin{equation}
  \label{eq:22}
  d_V \alpha (X_1, \dots, X_n) = d \alpha (\Pi_V^{\omega} X_1,
  \dots, \Pi_V^{\omega} X_n),
\end{equation}
where $\alpha$ is a $(n-1)$-form. The vertical forms $\Omega_V^*
(E)$ equipped with the vertical differential $d_V$ define the
\emph{vertical complex}.

Let's now suppose that it is possible to define a global section
$\sigma: X\rightarrow E$. This section defines a global
trivialization $\varphi_{\sigma}: X \times K \rightarrow E$, where
$\varphi_{\sigma} (x,g)=\sigma (x) \cdot g$. This trivialization
induces a distinguished connection $\omega_{\sigma}$ on $E$ such
that the pullback connection $\tilde{\omega}_{\sigma}\doteq
\varphi_{\sigma}^{\ast}\omega_{\sigma}$ coincides with the
canonical flat connection on $X \times K$
\cite{kobayashi63:_found}. Roughly speaking, the horizontal
distribution defined by $\omega_{\sigma}$ at $p=\sigma (x)$ is
tangent to the section $\sigma$. The vertical complex defined by
the connection $\tilde{\omega}_{\sigma}$ can be naturally
identified with the de Rham complex of $K$. This implies that
$d_{V}=d_{K}$. Since the connection form $\tilde{\omega}_{\sigma}$
is a $\liek$-valued $K$-invariant vertical form, it can be
identified with the Maurer-Cartan form $\theta_{MC}$ of the group
$K$. On the contrary, a general connection $\omega$ defines a
splitting of $TE$ which does not coincide with the splitting
induced by the section $\sigma$. In other words, the horizontal
distribution defined by $\omega$ is not tangent to $\sigma$. In
fact, the pullback connection $\varphi_{\sigma}^{\ast}\omega$ at
$(x,g)$ can be written as a sum
\begin{equation} \label{eq:59}
\varphi_{\sigma}^{\ast}\omega =ad_{g^{-1}}\varpi+\theta_{MC},
\end{equation}
where $\varpi=\sigma^{\ast} \omega \in \Omega^{1}\left(X
\right)\otimes \liek$ and $\theta_{MC}\in \liek^{*} \otimes \liek$
is the Maurer-Cartan form of $K$ \cite{Choquet-Bruhat}.

We will denote connections on the $G$-principal fiber bundle $P
\rightarrow M$ by the letter $A$. In a local trivialization $P|_U
\simeq U \times G$ we have an induced local $\liek$-valued 1-form
$A_U $. The forms $A_U$ are the so-called \emph{gauge
  fields}~\cite{faddeev91:_gauge}.\footnote{In a covariant framework,
  the connection $A$ can be regarded as the spatial part of a
  connection $\mathcal{A}$ on $\mathcal{P} \rightarrow \mathcal{M}$.
  In fact, since we can identify $\mathcal{P}$ with $\mathbb{R} \times
  P$ each connection $\mathcal{A}$ has a canonical decomposition
  $\mathcal{A} = A(t) + A_0 (t) dt$, where $A(t)$ is a time-evolving
  connection on $P$ and $A_0 (t)$ is a time-evolving section of $\ad =
  P \times_G \liea $. The action of the gauge group $\mr{G}$ on $A_0$
  is induced by the natural action on associated bundles. This action
  is the restriction of the action of the automorphism group of
  $\mathcal{P}$ to $\left\{t\right\}\times M$.} The configuration
space of all connections is an affine space modelled on the vector
space $\Omega^1(M, \liea)$ consisting of 1-forms with values on
the adjoint bundle $\ad $. The gauge group $\mr{G}$ acts on this
configuration space by affine transformations. We will fix a
metric $g$ on $M$ and an invariant scalar product $tr$ on $\liea$.
These data together with the corresponding Hodge operator $*$
induce a metric on $\Omega^1(M, \liea)$. Hence, a metric can be
defined in the spaces $\Omega^k(M, \liea), \; k \geq 1,$ by means
of the expression
\begin{equation}
  \label{eq:16}
  \langle \Omega_1, \, \Omega_2 \rangle = \int_M \, tr (\Omega_1 * \Omega_2).
\end{equation}

Since the action of $\mr{G}$ on the configuration space of
connections is not free, the quotient generally is not a manifold.
This problem can be solved by using \emph{framed connections}
\cite{donaldson90}. The letter $\mr{A}$ will denote the space of
\emph{framed connnections} of Sobolev class $L^2_{l-1}$ for the
metric defined in~\eqref{eq:16}, where $l $ is a fixed number
bigger than 2. The group $\mr{G} $ is the group of gauge
transformations of Sobolev class $L^2_l$. The action of $\mr{G}$
on $\mr{A}$ is free (see Ref.\cite[section 5.1.1]{donaldson90}).
We will denote by $\mr{B}$ the quotient $\mr{A} / \mr{G}$.
Uhlenbeck's Coulomb gauge fixing theorem (see Ref.\cite[section
2.3.3]{donaldson90}) implies, for a generic metric $g$, local
triviality. Hence $\mr{A} \rightarrow \mr{B}$ is a
$\mr{G}$-principal bundle.

The initial geometric arena for our construction is the pullback
$G$-principal bundle $p^*(P) \rightarrow \mr{A} \times M$, which
is obtained by taking the pullback of the bundle $P\rightarrow M$
by the projection $\mr{A} \times M \xrightarrow{p} M$. The bundle
$p^*(P)$ can be identified with $\mr{A} \times P$:
\begin{equation*}
 \begin{xy}
   \xymatrix{
     p^* (P) = \mr{A} \times P \ar[d] \ar[r] & P \ar[d] \\
     \mr{A} \times M \ar[r]^p & M.
     }
 \end{xy}
\end{equation*}

The gauge group $\mr{G}$ has an action on $\mr{A} \times M$
induced by its action on $\mr{A}$. This action is covered by the
action of $\mr{G}$ on $p^*(P)$ induced by its action on both
$\mr{A}$ and $P$.
\begin{prop}
 The bundle $p^*(P)\rightarrow \mr{A} \times M$ induces a $G$-principal
bundle~\cite{atiyah84:_dirac}
$$
   Q = (\mr{A} \times P) / \mr{G} \xrightarrow{\rho}
\mr{B} \times M.
$$
\end{prop}
We therefore have the following tower of principal bundles: $$
 \begin{xy}
 \xymatrix{
  \mr{G} \ar[r]  & p^*(P) = \mr{A} \times P \ar[d]^q \\ G \ar[r]
&
  Q = (\mr{A} \times P) / \mr{G} \ar[d]^{\rho} \\
   & \mr{B} \times M.
}
\end{xy}
$$

The fiber bundle $p^*(P)= \mr{A} \times P \rightarrow \mr{A}
\times M$ can be considered a universal family (parameterized by
the space $\mr{A}$) of fiber bundles $P \xrightarrow{\pi} M$ with
tautological connections $A$. The universal family
$\mathbf{A}^{U}$ of tautological connections is defined in the
following way. Let $(A, x)$ be a point of $\mr{A} \times M$. Then
the elements of the fiber of $\mr{A} \times P$ over $(A, x)$ have
the form $(A, \, p)$, with $p \in \pi^{-1}(x)$. Let us fix one of
these elements. Let $v \in T(\mr{A} \times P)_{(A, \, p)}$ be a
tangent vector such that $\pi_* (v) \in TM \subset T(\mr{A} \times
M)$ (i.e., $v$ is tangent to a copy of $P$ in $\mr{A} \times P$).
Then $\mathbf{A}^{U} (v) = A(v)$. We will write $H^U$ for the
distribution associated to the family $\mathbf{A}^{U}$. For each
element $A \in \mr{A}$, the distribution $H^U$ induces the
distribution $H_A$ on $TP$ defined by the connection $A$. The
universal family of connections $\mathbf{A}^{U}$ allows us to
define parallel transports along paths contained in any copy of
$M$ inside $\mr{A} \times M$.

\vspace{0.5 cm}

Let's now pick a connection in the $\mr{G}$-principal bundle
$\mr{A} \rightarrow \mr{B}$. This connection will be defined by
means of a 1-form $\eta \in \Omega^1 (\mr{A}) \otimes \lieG$. We
will denote by $\mathcal{H}_{\eta}$ the corresponding equivariant
distribution. This connection will define a generalized notion of
gauge fixing.\footnote{In Ref.\cite{Narasimhan79} the authors
analyze the particular case of the Coulomb connection for a
$SU(2)$ Yang-Mills theory on $S^{3}\times\mathbb{R}$. The authors
point out that in the absence of a global section, the gauge can
be consistently fixed by means of such a connection.} In fact,
let's suppose that it is possible to define a global gauge fixing
section $\sigma: \mr{B} \rightarrow \mr{A}$. This section defines
a global trivialization $\varphi_{\sigma}: \mr{B} \times \mr{G}
\rightarrow \mr{A}$. One can then define an induced flat
connection $\eta_{\sigma}$ on $\mr{A}\rightarrow \mr{B}$ such that
the corresponding distribution $\mathcal{H}_{\eta_{\sigma}}$ is
always tangent to $\sigma$. In other words, the pullback
$\varphi_{\sigma}^{\ast}\eta_{\sigma}$ coincides with the
canonical flat connection on $\mr{B} \times \mr{G}$ (see
Ref.\cite{kobayashi63:_found}). This shows that a global gauge
fixing section $\sigma$ can always be expressed in terms of a flat
connection $\eta_{\sigma}$. On the contrary, a connection $\eta$
can not in general be integrated to a section. A local obstruction
is the curvature and a global one the monodromy. Besides, it is
always possible to define a \emph{global} connection $\eta$, even
when the topology of the fiber bundle $\mr{A} \rightarrow \mr{B}$
is not trivial (Gribov's obstruction). In
sections~\ref{sec:quant-hamilt-sett} and
~\ref{sec:path-integr-quant} we will use the connection $\eta$ as
a generalized gauge fixing for the path integral quantization of
Yang-Mills theory. In particular, we will show that the connection
$\eta$ does not have to be flat in order to induce a well defined
gauge fixing.

Let's consider now a particular example of a gauge fixing
connection. Due to the affine structure of $\mr{A}$ there is a
canonical diffeomorphism $T
  \mr{A} \simeq \mr{A} \times \Omega^1 (M, \liea)$. We will consider
  $\Omega^1 (M, \liea)$ with the inner product defined
  by~\eqref{eq:16}. Since $\ad$ is a vector bundle associated to the
  principal bundle $P$, any connection $A$ on $P$ induces a covariant
  derivative
  \begin{equation*}
    \label{eq:ex-17}
    d_A : \Omega^k (M, \liea) \rightarrow \Omega^{k+1} (M, \liea).
  \end{equation*}
  Let
  \begin{equation*}
    \label{eq:18}
    d_A^*: \Omega^{k+1} (M, \liea) \rightarrow \Omega^{k} (M, \liea)
  \end{equation*}
  be the adjoint operator. This is a differential operator of first
  order. Define
  \begin{equation}
    \label{eq:17}
    \mathcal{H}_A: \text{Ker} \{ d_A^*: \Omega^{1} (M, \liea)
    \rightarrow \Omega^{0} (M, \liea) \}.
  \end{equation}
Then $\mathcal{H}_A$ defines a connection on $\mr{A}$ called the
\emph{Coulomb connection} (see for example Ref.\cite[p.
56]{donaldson90}).

\vspace{0.5 cm}

The distribution $\mathcal{H}_{\eta}$ together with $H^U$ define a
smooth distribution $\widetilde{\mr{H}}$ on $Tp^*(P)$. If $(A, \,
p) \in p^*(P)$, then
$$
 \widetilde{\mr{H}}_{(A, \, p)} = \mathcal{H}_{\eta}(A) \oplus H^U (A) (p).
$$

\begin{prop}\label{transv}
 The distribution $\widetilde{\mr{H}}$ is transversal to the orbits
 of the action of $\mr{G}$ on $\mr{A} \times P$ and to the fibers of
$p^*(P) \rightarrow \mr{A} \times P$.
\end{prop}
\begin{proof}
 Let $(A, \, p) \in p^*(P)$. Then we have two homomorphisms of vector
 spaces $\iota : \liea \rightarrow T_p P$ and $\kappa: \lieG
 \rightarrow T_A \mr{A} $. These homomorphisms are induced by the
 principal bundle structures of $P$ and $\mr{A}$.  For each point $p$
 there is also a homomorphism of Lie algebras $\tau_p: \lieG
 \rightarrow \liea $ given by $\tau_p (\boldsymbol{\mathfrak{g}}) =
 \boldsymbol{\mathfrak{g}}(p)$,
 where we use the identification of the elements of $\lieG $ with
 equivariant maps $\boldsymbol{\mathfrak{g}}: P \rightarrow \liea$.
 With these definitions the tangent space to the orbit $F_{\mr{G}}$ of the
 action of $\mr{G}$ at the point $(A, \, p)$ is equal to
\begin{equation}
 \label{eq:t-orbit}
 TF_{\mr{G}}(A, \, p) = \left\{ \mathbf{v} - v \in T_A \mr{A} \oplus T_p P \;
\vert \; \mathbf{v}
 = \kappa (\boldsymbol{\mathfrak{g}}) \, \text{and} \, v = \iota (\tau_p
 (\boldsymbol{\mathfrak{g}})) \right\}.
\end{equation}

The tangent spaces to the orbits are contained in the sum of the
tangent spaces to the orbits in $\mr{A}$ and $P$. The proposition
follows from the fact that the connections are transversal to
these spaces.
\end{proof}
\begin{rem}
 The distribution $\widetilde{\mr{H}}$ does not define a connection
 on $p^* (P) \rightarrow
\mr{A} \times M$ due to the fact that the tangent space at $(A, \,
p)$
 has a decomposition
 \begin{equation*}
   Tp^* (P)_{(A, \, p)} = TF_{\mr{G}}(A, \, p) \oplus
   \underbrace{\mathcal{H}_{\eta}(A) \oplus H^U (A)
     (p)}_{\widetilde{\mr{H}}_{(A, \, p)}} \oplus \iota(\liea),
 \end{equation*}
 being $\iota(\liea)$ the vertical subspace. Hence, the distribution
 $\widetilde{\mr{H}}$ does not define vector spaces complementary
 to the vertical subspace $\iota(\liea)$.
 However there is a reason for the introduction of this distribution
 which is explained by the following Lemma.
\end{rem}
\begin{lema}
 The distribution $\widetilde{\mr{H}}$ is $\mr{G}$-invariant and
 induces a connection $\mr{H}$ on $Q= (\mr{A} \times P) / \mr{G} \rightarrow
\mr{B} \times M $.
\end{lema}

This lemma follows from the invariance of the distributions
$\mathcal{H}_{\eta}$ and $H_{U}$.

\vspace{0.5 cm}

Let $\mr{E}$ be the connection on the bundle $p^* (P) \rightarrow
\mr{A} \times M$ obtained as the pullback of the connection
$\mr{H}$ by the projection $\mr{A} \times M \rightarrow \mr{B}
\times M$. This pullback can be understood either in the language
of distributions or in the language of forms. Let's consider the
diagram
\begin{displaymath}
   \xymatrix{
     & p^* (P) = \mr{A}\times P \ar[dr] \ar[dl]_{q} & \\
     Q=\frac{\mr{A}\times P}{\mr{G}} \ar[dr] & & \mr{A}\times M \ar[dl] \\
     & \mr{B} \times M. &
     }
\end{displaymath}

The map $q: p^* (P) \rightarrow Q $ induces a map $q_* : Tp^* (P)
\rightarrow TQ $. At each point $(A, \, p)$ the subspace of
$T_{(A, \, p)} p^* (P) $ which defines $\mr{E}$ is $q_*^{-1} (
\mr{H}_{q(A, \, p)})$. The $\liea$-valued 1-form $\AA \in
\Omega^{1}\left(\mr{A} \times P\right)\otimes \liea$ associated to
$\mr{E}$ is the pullback by $q$ of the 1-form associated to
$\mr{H}$. We will now identify this distribution and this 1-form.

The distribution which defines the new connection at each point
$(A, p)$ is the direct sum
\begin{equation}\label{ee1}
\mr{E}_{\left(A, p \right)} = TF_{\mr{G}}(A, \, p) \oplus
\underbrace{\mathcal{H}_{\eta}(A) \oplus H^U (A)
(p)}_{\widetilde{\mr{H}}_{(A, \, p)}}.
\end{equation}

If $(A, p ) \in \mr{A} \times P$ and $v = v_1 + v_2 \in T_A \mr{A}
\oplus T_p P$, then we define a $\liea$-valued 1-form $\AA$ on
$\mr{A} \times P$ given by
\begin{equation}
  \label{eq:240}
  \AA (v) = \mathbf{A}^{U}_{(A, \, p)} (v_2) + \eta (v_1) (p).
\end{equation}

We will now show that the horizontal distribution defined by $\AA$
is effectively given by (\ref{ee1}).

\begin{lema}\label{vert}
  If $ v \in TF_{(A, \, p)} \oplus \widetilde{\mr{H}}_{(A, \, p)}$,
then $\AA (v) = 0$.
\end{lema}
\begin{proof}

$(i)$ If $v \in H^U (A) (p) \subset \widetilde{\mr{H}}_{(A, \,
p)}$, then $$\AA (v) = \mathbf{A}^{U}_{(A, \, p)} (v) = A (v) =
0$$ by definition of the connection $A$.

$(ii)$ If $v \in \mathcal{H}_{\eta}(A) \subset
\widetilde{\mr{H}}_{(A, \, p)}$, then $$\AA (v) = \eta (v) (p) =
0$$ by definition of the connection $\eta$.

$(iii)$ If $v= \kappa (\boldsymbol{\mathfrak{g}}) - \iota (\tau_p
(\boldsymbol{\mathfrak{g}})) \in TF_{(A, \, p)}$, then
\begin{eqnarray*}
\AA \left( v \right) &=& - \mathbf{A}^{U}_{(A, \, p)} (\iota
(\tau_p
 (\boldsymbol{\mathfrak{g}}))) + \eta \left(\kappa
   (\boldsymbol{\mathfrak{g}}) \right) (p)
\\ &=& -A
 (\iota (\tau_p (\boldsymbol{\mathfrak{g}}))) + \eta \left(\kappa
   (\boldsymbol{\mathfrak{g}})
 \right) (p)
\\ &=& - \tau_p (\boldsymbol{\mathfrak{g}}) + \boldsymbol{\mathfrak{g}} (p)
\\ &=& 0,
\end{eqnarray*}
where we have used that by definition of connection $A \circ \iota
= id_{\liea}$ and $\eta \circ \kappa = id_{\lieG}$.
\end{proof}

\begin{rem}
  It is the gauge fixing connection $\mathcal{H}_{\eta}$ that allows
  us to make the decompositions~\eqref{ee1} and~\eqref{eq:240} of the
  horizontal distribution $\mr{E}$ and the corresponding 1-form
  $\AA$. The reason is that these kinds of decompositions require the
  choice of a complement to a subspace of a vector space.
\end{rem}

\begin{rem}
An important difference with the work of Baulieu and Singer for
topological Yang-Mills theory is that in
Ref.\cite{Baulieu-Singer1988} the connection $\omega$ is a
\emph{natural} connection, which is defined by using the
\emph{orthogonal} complements to the orbits of $G$. In order to
define this orthogonal complements one uses the fact that the
space $\mr{A} \times P$ has a Riemannian metric invariant under
$\mr{G} \times G$ (see Ref.\cite{atiyah84:_dirac} for details). In
our case the connection $\AA$, being tautological in the factor
$P$, is not natural in the factor $\mr{A}$, in the sense that the
gauge fixing connection $\eta$ can be freely chosen. This freedom
is in fact the freedom to choose the gauge.
\end{rem}

\section{Extended curvature and the BRST complex}
\label{sec:3}

In this section we will begin to consider the rich geometric
structure induced by the connection $\mr{H}$. We will do this
through the pullback form $\AA$ and its curvature $\FF$. Since we
have a diffeomorphism $p^*(P) \simeq \mr{A} \times P$, the de Rham
complex of $\liea$-valued forms on $p^*(P)$ is the graded tensor
product of the de Rham complexes of $\mr{A} $ and $P$, i.e.
\begin{equation}
  \label{eq:3}
  \Omega^* (p^*(P)) \otimes \liea \simeq \Omega^* ( \mr{A}) \otimes
  \Omega^* (P) \otimes \liea.
\end{equation}
This fact has two consequences. Firstly, the forms we are
considering are naturally bigraded. Secondly, the exterior
derivative $\Delta$ in $p^*(P)$ can be decomposed as
$\Delta=\delta + d $, where $\delta$ and $d$ are the exterior
derivatives in $\mr{A}$ and $P$ respectively. Since the forms that
we are considering are equivariant forms, the right complex to
study these forms is
\begin{equation}
  \label{eq:12}
  \Omega^* ( \mr{A}) \otimes \left(
  \Omega^* (P) \otimes_G \liea \right).
\end{equation}

Since
\begin{equation}
  \label{eq:10000}
   \Omega^* ( \mr{A}) \otimes \left( \Omega^0 (P) \otimes_G \liea
   \right) \simeq  \Omega^* ( \mr{A}) \otimes \lieG,
\end{equation}
the $\lieG$-valued equivariant $k$-forms on $\mr{A}$ can be
considered as elements of bidegree $(k, 0)$ of the
complex~\eqref{eq:12}. In particular the connection form $\eta$
defines an element of bidegree $(1,0)$ of this complex.

Using the splitting of the exterior derivative and the
decomposition of forms we obtain the following decomposition of
the curvature $\mathbb{F}$:
$$
 \mathbb{F}=\Delta \mathbb{A}+\frac{1}{2}\left[
 \mathbb{A},\mathbb{A}\right] =%
\mathbb{F}^{\left( 2,0\right) }+\mathbb{F}^{\left(
   1,1\right) }+\mathbb{F}^{\left( 0,2\right) },
$$
where
\begin{eqnarray}
\mathbb{F}^{\left( 2,0\right) } &=&\delta \eta +\frac{1}{2}\left[
\eta ,\, \eta \right] \equiv \phi, \label{phi} \\
\mathbb{F}^{\left( 1,1\right) } &=&\delta \mathbf{A}^{U}+d\eta
+\left[
 \mathbf{A}^{U}, \, \eta
\right] \equiv \psi, \label{psi}   \\ \mathbb{F}^{\left(
0,2\right) } &=& d \mathbf{A^{U}}+\frac{1}{2}\left[
 \mathbf{A}^{U}, \, \mathbf{A}^{U}\right]
\equiv \mathbf{F}^{U}. \label{F}
\end{eqnarray}

The $\left(0,2\right)$-form $\mathbf{F}^{U}$ is the universal
family of curvature forms corresponding to the universal family of
connections $\mathbf{A}^{U}$. In this sense, equation~\eqref{F} is
the extension to families of the usual Cartan's structure
equation. The $\left(2,0\right)$-form $\phi$ is the curvature of
the connection $\eta$. The $\left( 1,1\right)$-form $\psi$ is a
mixed term which involves both the gauge fields and the gauge
fixing connection. This last term shows that this construction
mixes in a non-trivial way the geometric structures coming from
the fiber bundles $P \rightarrow M$ and $\mr{A} \rightarrow
\mr{B}$.

We will now decompose equations~\eqref{phi} and~\eqref{psi} in
order to recover the usual BRST transformations of the gauge and
ghost fields.  Let $\Omega_V^k (\mr{A})$ for $k > 0$ be the
vertical differential forms induced by $\eta$. We define
$\Omega_V^0 (\mr{A}) = \mathcal{C}^{\infty} (\mr{A} )$. The
differential of the de Rham complex of $\mr{A}$ induces a
differential $\delta_V$ on the vertical forms. The complex
$\Omega_V^* ( \mr{A}) \otimes \Omega^* (P) \otimes_G \liea $ will
be termed the \emph{vertical complex}.

This decomposition shows that we can identify the vertical complex
with a subcomplex of $\Omega^* (\mr{A}) \otimes \Omega^*(P)
\otimes_G \liea $. Via this identification, the $(1,0)$-form
$\eta$ can be identified with a vertical form. In particular, we
see that the $(1, 1)$-forms have a decomposition in two terms, one
of which is the part of degree $(1, 1)$ of the vertical complex.

We will now write the explicit decomposition of both sides of the
equation~\eqref{psi}. Let $\delta=\delta_{V} + \delta_{H}$ be the
decomposition of the de Rham differential on $\mr{A}$ induced by
$\eta$. On degree $(0, *)$ the decomposition is given by the
following definition. Let $p_V$ and $p_H$ be the projectors onto
the factors associated with the decomposition into vertical and
horizontal forms. If we think of elements of $\Omega^0(\mr{A})
\otimes \Omega^1(P) \otimes_G \liea$ as $\Omega^1(P) \otimes_G
\liea$-valued functions on $\mr{A}$, then $\delta$ has a natural
decomposition $\delta=\delta_{V} + \delta_{H}$, where $\delta_V =
p_V \circ \delta$ and $\delta_H = p_H \circ \delta$.

The universal family of connections $\mathbf{A}^{U}$ can be
interpreted as a function $\mathbf{A}^{U}: \mr{A} \rightarrow
\Omega^1 (P) \otimes_G \liea$. We have a splitting
\begin{equation*}
 \label{eq:sp1}
 \delta \mathbf{A}^{U} = \delta_V  \mathbf{A}^{U} + \delta_{H} \mathbf{A}^{U} \in
\Omega^1
 (\mr{A}) \otimes \Omega^1 (P) \otimes_G \liea ,
\end{equation*}
where $\delta_V \mathbf{A}^{U}$ is an element of the vertical
complex. The universal family of connections $\mathbf{A}^{U}$
induces a family of connections on each of the vector bundles
associated to $P$ and in particular on $\ad $. This family of
connections can be seen as a homomorphism
\begin{equation}
 \label{eq:da}
 d_{\mathbf{A}^{U}} : p^* (\lieG) \rightarrow p^* \left(\Omega^1 (P) \otimes
\liea\right),
\end{equation}
given on each copy  $\left\{ A \right\} \times \lieG $ by the
covariant derivative $d_A: \lieG \rightarrow \Omega^1 (P) \otimes
\lieG$ associated to the connection $A$. Recall that sections of
$\ad$ can be seen as equivariant functions $P \rightarrow \liea $
and that the covariant derivative is $d_A = d \circ \pi_A$, where
$d$ is the exterior derivative in $P$ and $\pi_A : TP \rightarrow
TP$ is the horizontal projection. These constructions are
equivariant, which explains the codomain in
equation~\eqref{eq:da}. The term $d_{\mathbf {A}^{U}} \eta $ is by
definition the extension
\begin{equation*}
 \label{eq:dd}
 \mathbf{1} \otimes d_{\mathbf{A}^{U}} : \Omega^1 (\mr{A})  \otimes
 \lieG = \Omega^1 (\mr{A}) \otimes (\Omega^0(P) \otimes_G \liea)  \rightarrow \Omega^1 (\mr{A})  \otimes \Omega^1 (P) \otimes_G \liea
\end{equation*}
applied to $\eta$. Since the homomorphism $\mathbf{1} \otimes
d_{\mathbf{A}^{U}} $ acts on the second factor, it preserves
vertical forms. It follows that $d_{\mathbf{A}^{U}} \eta = d \eta
+ [\mathbf{A}^{U}, \eta]$ is a vertical form. From these remarks
we see that the vertical summand of the left hand side of equation
\eqref{psi} is
\begin{equation}
 \label{eq:bbrrsstt}
 \delta_V \mathbf{A}^{U} + d \eta + [\mathbf {A}^{U}, \, \eta].
\end{equation}

We will consider now the right hand side of the equation
\eqref{psi}. We will demonstrate the following proposition.

\begin{prop}
  The element $\psi$ is horizontal for the connection $\eta$.
\end{prop}

\begin{proof}
Since the connection $\AA$ is the pullback of a connection on the
$G$-fiber bundle $Q = (\mr{A} \times P)/\mr{G} \rightarrow
\mr{A}/\mr{G} \times M$ the same is true for the curvature $\FF$.
If we denote $\omega_{\mr{H}}$ and $\mathcal{F}_{\mr{H}}$ for the
connection and curvature forms of the distribution $\mr{H}$ on
$Q$, then one has
\begin{eqnarray*}
\AA &=& q^{\ast}\omega_{\mr{H}},
\\ \FF&=&q^{\ast}\mathcal{F}_{\mr{H}},
\end{eqnarray*} where $q$ is the projection $\mr{A} \times P
\xrightarrow{q}  Q=(\mr{A} \times P) / \mr{G}$.

If $X$ is a vector tangent to the fibers $TF_{\mr{G}}$ of the
action of $\mr{G}$ on $\mr{A} \times P$ then the contraction
$\imath_{X} \FF$ is equal to  $\imath_{X}
q^{\ast}\mathcal{F}_{\mr{H}} = \imath_{q_{\ast}X}
\mathcal{F}_{\mr{H}}$. Since $X$ has the form $X= (\mathbf{v}, -v)
\in TF_{\mr{G}}$ with $TF_{\mr{G}}$ given by (\ref{eq:t-orbit}),
then $q_{\ast}X=0$. This results from the fact that the vectors
tangent to $\mr{G}$ are projected to zero when we take the
quotient by the action of $\mr{G}$. Hence, the contraction
$\imath_{X} \FF$ of the curvature $\FF$ with a vector $ X =
(\mathbf{v}, -v)$ tangent to the fibers given
by~\eqref{eq:t-orbit} is zero:

\begin{equation}\label{ff1}
\imath_{(\mathbf{v}, -v)} \FF = \imath_{q_{\ast}(\mathbf{v}, -v)}
\mathcal{F}_{\mr{H}} = 0.
\end{equation}

An analysis of the different components of $\FF$ shows that

$\bullet$ $\imath_{(\mathbf{v}, -v)} \FF^{(2,0)} =
\imath_{(\mathbf{v})} \FF^{(2,0)} = 0 $ since  $\FF^{(2,0)}$ is
induced by the connection $\eta$ and $\mathbf{v}$ is vertical for
this connection.

$\bullet$ $\imath_{(\mathbf{v}, -v)} \FF^{(0,2)} = \imath_{-v}
\FF^{(0, 2)} = 0 $ since  $\FF^{(0,2)}$ is induced by the
connection $A$ and $-v$ is tangent to the fibers of $p^* (P) \to
\mr{A} \times P$.

$\bullet$ $\imath_{-v} \FF^{(1,1)} = 0$ since $-v$ is tangent to
the fibers of $p^* (P)\to \mr{A} \times P$.

From these remarks and equation~\eqref{ff1} it follows that
\begin{equation*}
 \label{eq:great}
 0 = \imath_{(\mathbf{v}, -v)} \FF = \imath_{(\mathbf{v}, -v)}
 \FF^{(1,1)} = \imath_{\mathbf{v}} \FF^{(1,1)} = \imath_{\mathbf{v}}
\psi.
\end{equation*}
\end{proof}

The equation~\eqref{psi} then splits in the equations
\begin{eqnarray}
\delta_V \mathbf{A}^{U} &=& - d_{\mathbf A^{U}} \eta,
\label{brstA}
\\ \delta_H \mathbf{A}^{U} &=& \psi.
\end{eqnarray}

\vspace{0.5 cm}

The equation for the (2, 0)-form $\phi$ can also be canonically
decomposed in components belonging to the vertical and horizontal
complexes. The differential $\delta$ acting on elements of
$\Omega^1(\mr{A}) \otimes \lieG = \Omega^1(\mr{A}) \otimes
\Omega^0(P) \otimes_G \liea$ also has a decomposition $\delta =
\delta_V + \delta_H$, where $\delta_V = \delta \circ p_V$ and
$\delta_H = \delta \circ p_H$. The horizontal part $\delta_H$
corresponds to the covariant derivative with respect to the
connection $\eta$. Therefore we have a splitting
\begin{eqnarray*}
  \delta \eta = \delta_V \eta + \delta_{H} \eta \in
  \Omega^2    (\mr{A}) \otimes  (\Omega^0(P) \otimes_G \liea) =
  \Omega^2    (\mr{A}) \otimes  \lieG.
\end{eqnarray*}

By definition of the curvature $\phi$ associated to the connection
$\eta$ we have
\begin{equation}
 \label{eq:brs1}
 \delta_{H} \eta = \phi.
\end{equation}

The vertical component of the equation is such that $\delta_V \eta
+ \delta_{H} \eta = \delta \eta= \phi - \frac{1}{2}\left[\eta,
\eta \right]$. We have then
\begin{equation}
 \label{eq:brs2}
   \delta_V \eta = -\frac{1}{2}\left[\eta, \eta \right].
\end{equation}

In the next section we will show how the BRST transformations of
the gauge and ghost fields can be obtained from equations
\eqref{brstA} and \eqref{eq:brs2} respectively.

\section{The relationship between the gauge fixing connection and the ghost field}
\label{sec:4}

The proposed formalism allows us to further clarify the
relationship between the gauge fixing and the ghost field. To do
so, we shall first work in a local trivialization
$\varphi_{\sigma_{i}}:U_{i} \times \mr{G} \rightarrow
\pi^{-1}\left(U_{i} \right)$ defined by a local gauge fixing
section $\sigma_{i}: U_{i} \rightarrow \mr{A}$ over an open subset
$U_{i}\subset \mr{A}/\mr{G}$. Let $\tilde{\eta}$ be the pull-back
by $\varphi_{\sigma_{i}}$ of the connection form $\eta$ restricted
to $\pi^{-1}\left(U_{i}\right)$. As we have seen in
section~\ref{sec:2}, the connection form $\tilde{\eta}$ at $([A],
\mathbf{g})\in U_{i} \times \mr{G}$ takes the form
\begin{equation}\label{eq:csi}
\tilde{\eta}=ad_{\mathbf{g}^{-1}}\eta_{i}+\theta_{MC},
\end{equation}
where $\eta_{i}=\sigma_{i}^{\ast} \eta \in \Omega^{1}\left(U_{i}
\right)\otimes \mathcal{L}ie \left(\mr{G}\right)$ is the local
form of the connection $\eta$ and $\theta_{MC}\in \mathcal{L}ie
\left(\mr{G}\right)^{*} \otimes \mathcal{L}ie \left(\mr{G}\right)$
is the Maurer-Cartan form of the gauge group $\mr{G}$
\cite{Choquet-Bruhat}. The Maurer-Cartan form satisfies the
equation
\begin{equation}\label{eq:chu}
\delta_{\mr{G}}\theta_{MC}=-\frac{1}{2}\left[\theta_{MC},
\theta_{MC}\right].
\end{equation}

The formal resemblance between this equation and the BRST
transformation of the ghost field $c$ led in Ref.\cite{bonora83}
to the identification of $\delta_{BRST}$ and $c$ with the
differential $\delta_{\mr{G}}$ and the Maurer-Cartan form
$\theta_{MC}$ of $\mr{G}$ respectively. Hence, equation
\eqref{eq:csi} shows that the ghost field can be identified with
the canonical vertical part of the gauge fixing connection $\eta$
expressed in a local trivialization.

We will now show that it is possible to recover the standard BRST
transformation of the gauge field $\delta_{BRST} A = - d_{A} c$
from equation \eqref{brstA}. To do so, we will first suppose that
it is possible to define a global gauge fixing section $\sigma:
\mr{B} \rightarrow \mr{A}$. As we have seen in
section~\ref{sec:2}, the associated trivialization
$\varphi_{\sigma}$ induces a distinguished connection
$\eta_{\sigma}$ such that
$\varphi_{\sigma}^{\ast}\eta_{\sigma}=\theta_{MC}$. Therefore,
equation \eqref{brstA} yields in this trivialization
\begin{equation}\label{eq:xih}
\delta_{\mr{G}} \mathbf{A}^{U} = - d_{\mathbf{A}^{U}}\theta_{MC}.
\end{equation}

This equation is an extension to families of the usual BRST
transformation of the gauge field $A$. If a global gauge fixing
section cannot be defined, then it is possible to show that the
usual BRST transformation of $A$ is valid locally. In fact, since
$\delta_V \mathbf{A}^{U}$ is a vertical form, the substitution of
the local decomposition \eqref{eq:csi} in equation \eqref{brstA}
yields the BRST transformation \eqref{eq:xih}. We can thus
conclude that the usual BRST transformation of $A$ given by
\eqref{eq:xih} is only valid in a local trivialization of $\mr{A}
\rightarrow \mr{A}/\mr{G}$. Therefore, equation \eqref{brstA} can
be considered the globally valid BRST transformation of the gauge
field $A$. In fact, we will now show that equation \eqref{brstA}
plays the same role as the usual BRST transformation of the gauge
field. To do so, we have to take into account that the BRST
transformation $\delta_{BRST} A = - d_{A} c$ defines a
\emph{general} infinitesimal gauge transformation of $A$
\cite{bonora83}. In order to obtain a particular gauge
transformation from this general expression, it is necessary to
choose an element $\xi \in \mathcal{L}ie \left(\mr{G}\right)$. In
doing so, the usual gauge transformation of $A$ is recovered
\begin{equation}\label{eq:gta}
\delta A = (\delta_{BRST} A)(\xi)=- d_{A} (c (\xi))=- d_{A} \xi.
\end{equation}

Let's now consider equation \eqref{brstA}. According to the
definition of connections, $\eta (\xi^{\sharp})=\xi$, where
$\xi^{\sharp}$ is the fundamental vector field in $T\mr{A}$
corresponding to $\xi \in \mathcal{L}ie \left(\mr{G}\right)$.
Therefore, equation \eqref{brstA} yields
$$\delta \mathbf{A}^{U} =(\delta_V \mathbf{A}^{U})(\xi^{\sharp})= -
d_{\mathbf A^{U}}(\eta (\xi^{\sharp}))=- d_{\mathbf A^{U}}\xi.$$

This equation is the universal family of infinitesimal gauge
transformations defined in \eqref{eq:gta}. Therefore, equation
\eqref{brstA} can be consistently considered the globally valid
extension to families of the usual BRST transformation of $A$.

The identification of the ghost field with the canonical vertical
part of the gauge fixing connection $\eta$ depends on a particular
trivialization of the fiber bundle. Nevertheless, we will now show
that it is also possible to understand the relationship between
$c$ and $\eta$ without using such a trivialization. To do so it is
necessary to introduce the \emph{Weil algebra} of the gauge group
$\mr{G}$ (see Refs.\cite{dubois}, \cite{guillemin99},
\cite{szabo00}). The Weil algebra of a Lie algebra $\liek$ is the
tensor product $\mathcal{W}\left(\liek \right)= S^* \liek^{*}
\otimes \bigwedge^* \liek^{*}$ of the symmetric algebra $S^*
\liek^{*}$ and the exterior algebra $\bigwedge^* \liek^* $ of
$\liek^*$ (where $\liek$ and $\liek^{*}$ are dual spaces). Let
$T_{a}$ and $\vartheta^{a}$ be a base of $\liek$ and $\liek^{*}$
respectively. The Weil algebra is then generated by the elements
$\theta^{a}= 1 \otimes \vartheta^{a} $ and
$\zeta^{a}=\vartheta^{a} \otimes 1$. The graduation is defined by
assigning degree 1 to  $\theta^{a} $ and degree 2 to $\zeta^{a}$.
Let's define the elements $\theta$ and $\zeta$ in
$\mathcal{W}\left(\liek \right) \otimes \liek$ as
$\theta=\theta^{a} \otimes T_{a} $ and $\zeta = \zeta^{a} \otimes
T_{a}$ respectively. In fact, the element $\theta$ is the
Maurer-Cartan form $\theta_{MC}$ of the group. The Weil's
differential $\delta_{\mathcal{W}}$ acts on these elements by
means of the expressions
\begin{eqnarray*}
\delta_{\mathcal{W}}\theta &=& \zeta - \frac{1}{2}\left[\theta,
\theta \right],
\\ \delta_{\mathcal{W}}\zeta &=& -\left[\theta, \zeta
\right].
\end{eqnarray*}

These equations reproduce in the Weil algebra the Cartan's
structure equation and the Bianchi identity respectively. What is
important to note here is that the connection $\eta$ in the
$\mr{G}$-principal bundle $\mr{A} \rightarrow \mr{A}/\mr{G}$ can
be defined as the image of $\theta_{MC}$ under a particular
\emph{Chern-Weil homomorphism}
\begin{eqnarray*}\omega: \left(\mathcal{W}\left(\lieG \right)\otimes \lieG,
\delta_{\mathcal{W}}\right) &\longrightarrow&
\left(\Omega^{\ast}\left(\mr{A}\right)\otimes \lieG,\delta\right)
\\ \theta_{MC} &\mapsto& \eta.
\end{eqnarray*}

Therefore, the Weil algebra is a universal model for the algebra
of a connection and its curvature. In this way, a gauge fixing of
the theory by means of a connection $\eta$ can be defined by
choosing a particular Chern-Weil's immersion $\omega$ of the
``\emph{universal connection}'' $\theta_{MC}$ into $\Omega^{\ast}
(\mr{A}) \otimes \lieG$. Hence, the ghost field can be considered
a universal connection whose different immersions $\omega$ define
different gauge fixings of the theory.

We will now show that the usual BRST transformation of the ghost
field can be recovered from equation \eqref{eq:brs2}. To do so, it
is necessary to restrict the attention to the vertical complexes.
Indeed, the connection $\eta \in
\Omega^{1}\left(\mr{A}\right)\otimes \lieG$ defines a homomorphism
of differential algebras
$$\omega_{V}: (\lieG^{\ast}\otimes\lieG, \delta_{\mr{G}}) \rightarrow
(\Omega^{\ast}_{V} (\mr{A}) \otimes \lieG,\delta_{V}),$$ with
$\omega_{V}(\theta_{MC})=\eta$. This means that
$\omega_{V}(\delta_{\mr{G}}\alpha)=\delta_{V}(\omega_{V}(\alpha))$
for $\alpha\in \lieG^{\ast}\otimes \lieG$. Therefore, equation
\eqref{eq:brs2} yields
\begin{eqnarray*}
\delta_V (\omega_{V}(\theta_{MC})) &=&
-\frac{1}{2}\left[\omega_{V}(\theta_{MC}), \omega_{V}(\theta_{MC})
\right]
\\ \omega_{V}(\delta_{\mr{G}}\theta_{MC}) &=& \omega_{V}(-\frac{1}{2}\left[\theta_{MC}, \theta_{MC}
\right]).
\end{eqnarray*}

Therefore
$$\delta_{\mr{G}}\theta_{MC}=-\frac{1}{2}\left[\theta_{MC}, \theta_{MC}
\right],$$ which coincides with the ghost field's BRST
transformation.

\begin{rem}
Contrary to what is commonly done in order to reobtain the BRST
transformations for the ordinary (non-topological) Yang-Mills
case, it has not been necessary to impose the \emph{horizontality
conditions} $\phi=\psi=0$ on the extended curvature $\mathbb{F}$
(see for example Refs.\cite{Baulieu-Bellon},
\cite{Baulieu-Mieg1982}, \cite{Mieg}).
\end{rem}

\section{Path integral gauge fixing}
\label{sec:quant-hamilt-sett}

\subsection{Usual gauge fixing}

The central problem in the quantization of Yang-Mills theory is
computing the transition amplitudes
\begin{equation}
  \label{eq:1}
  \langle [A_0] \;  \vert \; [A_1] \rangle = \int_{T^* \mathcal{P}([A_0], [A_1])}
  \, \exp\{i S\} \; \mr{DA} \mr{D}\pi,
\end{equation}
where $S$ is the canonical action, $\mr{D}A$ is the Feynman
measure on the space of paths
\begin{equation*}
  \label{eq:2}
  \mathcal{P}([A_0], [A_1]) = \{ \gamma: [0, 1] \rightarrow \mr{A} /
  \mr{G} \, \vert \,
\gamma(i) = [A_i], \, i =0,1 \},
\end{equation*}
in $\mr{A} / \mr{G}$ and $\mr{D}\pi$ is a Feynmann measure in the
space of moments. The canonical Yang-Mills's action is given by
the expression
\begin{equation} \label{eq:ham}
S = \int dt \int d^{3}x
 \left(\dot{A}^{a}_{k}\pi^{k}_{a}-\mathcal{H}_{0}\left(\pi^{k}_{a},
B_{a}^{k}\right)-A^{a}_{0}\phi_{a}\right),
\end{equation}
with $\pi^{k}_{a}=F_{a}^{k0}$ and $B^{a}_{k}=\frac{1}{2}
\varepsilon_{kmn}F^{amn}$ (where $F^{a}_{mn}$ are the field
strengths).  The Yang-Mills Hamiltonian
$\mathcal{H}_{0}\left(\pi^{k}_{a}, B_{a}^{k}\right)$ is
\begin{equation}
\mathcal{H}_{0}\left(\pi^{k}_{a}, B_{a}^{k}\right)=
\frac{1}{2}\left[\pi^{k}_{a}\pi_{k}^{a}+ B_{a}^{k}B^{a}_{k}
\right],
\end{equation}
and the functions $\phi_{a}$ are
\begin{equation} \label{eq:gau}
\phi_{a}=-\partial_{k}\pi^{k}_{a}+f^{c}_{ab}\pi^{k}_{c}A^{b}_{k}.
\end{equation}

The pairs $(A^{a}_{k},\pi^{k}_{a})$ are the canonical variables of
the theory. The temporal component $A^{a}_{0}$ is not a dynamical
variable, but the Lagrange multiplier for the generalized Gauss
constraint $\phi_{a}\approx 0.$

The geometry of the quotient space $\mr{A}/ \mr{G}$ is generally
quite complicated. The usual approach is to replace the
integral~\eqref{eq:1} with an integral over the space of paths in
the affine space $\mr{A}$. To do so, one must pick two elements
$A_i \in\pi^{-1}[A_i]$ in the fibres $[A_i]\in \mr{A}/ \mr{G}$ ($i
=0,1$). Then one replaces the integral~\eqref{eq:1} by
\begin{equation}
  \label{eq:20}
  \langle A_0 \; \vert \; A_1 \rangle = \int_{T^*\mathcal{P}(A_0, \, A_1)}
  \, \exp\{i S\} \da \mr{D}\pi,
\end{equation}
where the integral is now defined on the cotangent bundle of the
following space of paths in $\mr{A}$
\begin{equation*}
  \label{eq:25}
   \mathcal{P}(A_0, \, A_1) = \{ \gamma: [0, 1] \rightarrow \mr{A}  \,
   \vert \, \gamma(0) = A_0, \, \gamma(1) = A_1 \}.
\end{equation*}

The problem with this approach is that it introduces an infinite
volume in the path integral, which corresponds to the integration
over unphysical degrees of freedom. The projection $\pi: \mr{A}
\rightarrow \mr{A} / \mr{G}$ induces a projection
$$
\widetilde{\boldsymbol{\pi}}: \mathcal{P} (A_0, \; A_1)
\rightarrow \mathcal{P}([A_0], \, [A_1]).
$$
The path group
$$
\widetilde{\mr{P}\mr{G}} = \{g(t):[0, \, 1] \rightarrow \mr{G} \,
\vert \, g(0) = g(1) = id_{\mr{G}} \}
$$
acts on $\mathcal{P} (A_0, \, A_1)$ by pointwise multiplication
and the fibers of $\widetilde{\boldsymbol{\pi}}$ consist of the
orbits of the action of $\widetilde{\mr{P}\mr{G}}$. Since the
action $S$ is invariant under the action of
$\widetilde{\mr{P}\mr{G}}$, one needs to extract the volume of
this group from the integral (\ref{eq:20}). In order to get rid of
this infinite volume the usual approach is to fix the gauge by
defining a section $\sigma: \mr{A} / \mr{G} \rightarrow \mr{A}$
such that $\sigma([A_i]) = A_i, \, i=0, 1$. The gauge fixing
section $\sigma$ induces a map $\bs{\tilde{\sigma}}$
$$
\begin{xy}
 \xymatrix{
   \mathcal{P}(A_0, \, A_1)  \ar[r]^{\widetilde{\boldsymbol{\pi}}}
   & \mathcal{P} ([A_0], \, [A_1]) \ar@/_1.2pc/[l]_{\bs{\tilde{\sigma}}},}
\end{xy}
$$
defined by $\bs{\tilde{\sigma}} (\gamma) = \sigma \circ \gamma$,
which is a section of $\widetilde{\boldsymbol{\pi}}$. This section
$\bs{\tilde{\sigma}}$ induces a trivialization
\begin{equation*}
  \label{eq:4}
  \mathcal{P}(A_0, \, A_1) \simeq \mathcal{P} ([A_0],
\, [A_1]) \times \widetilde{\mr{P}\mr{G}}
\end{equation*}
and a similar decomposition at the level of cotangent bundles. The
ghost field appears when one computes the Jacobian which relates
the corresponding measures. One can then use Fubini's theorem in
order to extract the irrelevant and problematic factor.

It is worth noting that the definition of a gauge fixing section
$\sigma$ in the $\mr{G}$-principal fiber bundle of fields $\mr{A}
\rightarrow \mr{A}/ \mr{G} $ is only an auxiliary step for
defining a section $\bs{\tilde{\sigma}}$ of the
$\widetilde{\mr{P}\mr{G}}$-projection $\mathcal{P} (A_0, A_1)
\xrightarrow{\widetilde{\boldsymbol{\pi}}} \mathcal{P}([A_0], \,
[A_1])$ in the space of paths \emph{where the path integral is
actually defined}.

\subsection{Generalized gauge fixing}

We will now consider in which sense the connection $\eta$ can be
used to fix the gauge. This gauge fixing will be globally
well-defined, even if there is a Gribov's obstruction. We shall
begin by considering paths in $\mr{A}$ such that the initial
condition $A_0$ is fixed and the final condition is defined
\emph{only up to a gauge transformation} (see Ref.\cite[p.
123]{Narasimhan79}). This means that the final condition can be
any element of the final fiber $\pi^{-1} [A_1]$. The corresponding
space of paths is
\begin{equation*}
  \label{eq:5}
  \mathcal{P} (A_0, \pi^{-1} [A_1]) = \{ \gamma:[0, \, 1] \rightarrow
  \mr{A}\, \vert   \, \gamma(0) = A_0, \, \pi(\gamma(1)) = [A_1] \}.
\end{equation*}

The relevant path group is now
\begin{equation*}
  \label{eq:6}
 \mr{P}\mr{G} = \{g(t):[0, \, 1] \rightarrow \mr{G} \, \vert \, g(0) =
  \text{id}_{\mr{G}} \}.
\end{equation*}

This group acts on $\mathcal{P} (A_0,  \pi^{-1} [A_1])$. This
actions defines the projection
\begin{equation*}
  \label{eq:7}
  \bs{\pi}:
\mathcal{P} (A_0,\pi^{-1} [A_1]) \rightarrow \mathcal{P} ([A_0],
\, [A_1]).
\end{equation*}

It is easy to show that the action of the path group
$\mr{P}\mr{G}$ on $\mathcal{P} (A_0, \pi^{-1} [A_1])$ is free. We
will not need to assume that it is a principal bundle.

The gauge fixing by means of the connection $\eta $ is defined by
taking parallel transports along paths in $\mr{A} / \mr{G}$ of the
initial condition $A_0 \in \pi^{-1} [A_0]$ (as has already been
suggested in Ref.\cite{Narasimhan79}). This procedure defines a
section $\bs{\sigma}_{\eta}$ of the projection $\bs{\pi}$
\begin{equation}
\label{eq:8}
\begin{xy}
 \xymatrix{
   \mathcal{P} (A_0, \pi^{-1} [A_1])  \ar[r]^{\bs{\pi}}
   & \mathcal{P}([A_0], [A_1]) \ar@/_1.2pc/[l]_{\bs{\sigma}_{\eta}}.}
\end{xy}
\end{equation}

The section $\bs{\sigma}_{\eta}$ sends each path
$\left[\gamma\right] \in \mathcal{P}([A_0], [A_1])$ to its
$\eta$-horizontal lift
$\gamma=\bs{\sigma}_{\eta}\left(\left[\gamma\right]\right) \in
\mathcal{P} (A_0, \pi^{-1} [A_1])$ starting at $A_0$.\footnote{The
theorem of  existence of parallel transport has been extended to
  infinite dimensions in Ref.~\cite[Theorem 39.1]{kriegl97}. It can be shown that under suitable assumptions the parallel transport depends
  smoothly on the path. We will therefore assume that the section
  $\bs{\sigma}_{\eta}$ is smooth and that its image is a smooth
  submanifold  of $\mathcal{P} (A_0,\pi^{-1} [A_1])$. By
  definition,
  this submanifold is transversal to the action of
  $\mr{P}\mr{G}$.}
A path $\gamma \in \mathcal{P} (A_0,\pi^{-1} [A_1])$ is in the
image of $\bs{\sigma}_{\eta}$ if and only if the tangent vectors
to $\gamma$ at each $A\in \mr{A}$ belong to the horizontal
subspaces $\mathcal{H}_{\eta}(A)$ defined by $\eta$. Recalling
that $\mathcal{H}_{\eta}(A) = \mbox{\emph{Ker} } \eta(A)$, this
local condition leads to the \emph{gauge fixing equation}
\begin{equation}
  \label{eq:15}
  \eta(\dot{\gamma}(t))=0, \quad \forall t.
\end{equation}

In local bundle coordinates this condition defines a non-linear
ordinary equation. The explicit form of this equation is given
in~\eqref{eq:final} (see Ref.\cite{Michor91} for details). In the
case of the Coulomb connection defined in \eqref{eq:17},
equation~\eqref{eq:15} becomes
\begin{equation}
  \label{eq:19}
  d_{\gamma(t)}^* \dot{\gamma}(t)=0,  \quad \forall t.
\end{equation}
This equation can be expressed in local terms on $M$ and $P$ for
each $t$.

If $\pi: X \rightarrow X/G$ is a quotient space by a principal
action, then any section $\sigma: X/G \rightarrow X$ induces a
global trivialization $\Phi : X/G \times G \xrightarrow{\simeq}
X$, where $\Phi([x], g) = \sigma([x])\cdot g$. It can be shown
that $\Phi$ is a diffeomorphism. It follows that the space of
paths $\mathcal{P} (A_0,\pi^{-1} [A_1])$ can be factorized as
$\mathcal{P} ([A_0], \, [A_1]) \times \mr{P}\mr{G}$. A similar
decomposition is induced at the level of cotangent bundles.

Strictly speaking the path integral is not an integral on the
space of fields $\mr{A}$, but rather an integral on the space of
paths. Hence, the section $\bs{\sigma}_{\eta}$ induced by the
connection $\eta$ suffices to get rid of the infinite volume of
the path group $\mr{P}\mr{G}$.

\begin{rem}
The generalized gauge fixing can be also be defined as the null
space of a certain functional as follows. The Lie algebra $\lieG$
of the gauge group $\mr{G}$ can be identified with the sections of
the adjoint bundle $\ad$. The invariant metric on $\liea$ induces
a metric $<\, , \, >_{\ad}$ on $\ad$. Using this metric we define
a $\mr{G}$-invariant metric on $\lieG = \Gamma (\ad)$ by
\begin{equation}
  \label{eq:314}
  \langle \sigma_1, \;  \sigma_2 \rangle_{\lieG}  =
  \int_M \, < \sigma_1(p), \, \sigma_2 (p) >_{\ad} \, dx.
\end{equation}

Then we define the functional $\mathcal{F}: \mathcal{P}
(A_0,\pi^{-1} [A_1]) \rightarrow \re $ by
\begin{equation}
  \label{eq:funcional}
  \mathcal{F} (\gamma) = \int_0^1 \, \lVert \eta (\dot{\gamma} (t))
  \rVert^2_{\lieG} \, dt
\end{equation}
\end{rem}

This is a positive functional and the image of the section
$\mathbf{\sigma}_{\eta}$ is the null space of $\mathcal{F}$.

\section{Faddeev-Popov method revisited}
\label{sec:path-integr-quant} We will now proceed to implement the
proposed generalized gauge fixing at the level of the path
integral. To do so, we will show that the usual Faddeev-Popov
method can also be used with this generalized gauge fixing. We
will then start by introducing our gauge fixing condition at the
level of the transition amplitude
\begin{equation} \label{kdu}
\langle A_0 \; \vert \; \pi^{-1}[A_1] \rangle =
\int_{T^*\mathcal{P}(A_0, \, \pi^{-1}[A_1])}
  \, \exp\{i S\} \da \mr{D}\pi.
\end{equation}

The first possible form of the gauge fixing condition is $ \delta
(\mathcal{F} (\gamma))$ where $\delta$ is the Dirac delta function
on $\re$ and $\mathcal{F} (\gamma)$ the functional
(\ref{eq:funcional}). This form is mathematically consistent and
does not require any product of distributions. This gauge fixing
condition has the direct exponential representation
\begin{eqnarray*}
\delta (\mathcal{F} (\gamma)) &=&\int d\lambda e^{i\lambda
\mathcal{F}(\gamma)}
\\ &=& \int d\lambda e^{i\lambda \int_{\gamma} \, \lVert \eta (\dot{\gamma} (t))
  \rVert^2_{\lieG} \, dt}
\\ &=& \int d\lambda e^{i\lambda \int_{\gamma}\int_{M} \, \lVert \eta (\dot{\gamma} (t))
  \rVert^2_{\liea} \, dx dt}
\end{eqnarray*}

The second form is based on the elementary observation that the
integral of a continuous positive function is zero if and only if
the function is zero at all points. One can then define the gauge
fixing condition
\begin{eqnarray}
\label{eq:9}
 \bs{\delta} (\eta (\dot{\gamma}))&=& \lim_{N}
\prod^{N}_{k = 1} \delta_{\lieG}(\eta (\dot{\gamma} (t_{k})))
\\ \nonumber &=& \lim_{N, M} \prod^{N}_{k = 1} \prod^{M}_{j =
1}\delta_{\liea}(\eta(\dot{\gamma}(t_{k}))(x_{j})),
\end{eqnarray}
where $\delta_{\lieG}$ is the delta function on $\lieG$ defined as
an infinite product of the Dirac delta $\delta_{\liea}$ on
$\liea$. If $T_a$ is a fixed basis of $\liea$, we can write
$\delta_{\liea}(\eta(\dot{\gamma}(t_{k}))(x_{j}))$ in terms of
$\delta (\eta(\dot{\gamma}(t_{k}))(x_{j})^a)$, where now the delta
function is the usual delta function on $\re^3$.

As usual we define the element $\Delta^{-1} [ \gamma ]$ as

  \begin{equation}\label{delta}
    \Delta^{-1} [ \gamma ] = \int_{\mr{PG}} \; \mr{D} g'
    \bs{\delta} (\eta
    (\dot{\gamma g'})),
  \end{equation}
where $\gamma g'$ denotes the right action of $\mr{PG}$ on
$\mathcal{P}(A_0, \, \pi^{-1}[A_1])$.

\begin{prop}
  The element $\Delta^{-1} [ \gamma ]$ is $\mr{PG}$ invariant.
\end{prop}
\begin{proof}
  \begin{eqnarray*}
    \label{eq:11}
    \Delta^{-1} [ \gamma g ] &=& \int_{\mr{PG}} \; \mr{D} g'
    \bs{\delta} (\eta
    (\dot{\gamma  g g'})) = \int_{\mr{PG}} \; \mr{D} (g g')
    \bs{\delta} (\eta
    (\dot{\gamma  g g'})) \\
    \nonumber &=& \int_{\mr{PG}} \; \mr{D} (g'')
    \bs{\delta} (\eta
    (\dot{\gamma g''})) = \Delta^{-1} [ \gamma ].
  \end{eqnarray*}
\end{proof}

We can then express the number one in the following way:
  \begin{equation}
    \label{eq:10}
    1 =  \Delta [ \gamma ] \int_{\mr{PG}} \; \mr{D} g'
   \bs{\delta} (\eta
    (\dot{\gamma  g'}))
  \end{equation}

Roughly speaking, the element $\Delta [\gamma]$ corresponds to the
determinant of the operator which measures the gauge fixing
condition's variation under infinitesimal gauge transformations.

It can be shown that the element $\Delta [\gamma]$ is never zero.
The local gauge fixing condition $\eta(\dot{\gamma}(t))=0$ induces
a well-defined section $\bs{\sigma}_{\eta}$ of the
$\mr{PG}$-projection $\bs{\pi}:\mathcal{P} (A_0, \pi^{-1} [A_1])
\rightarrow \mathcal{P}([A_0], [A_1])$ in the space of paths.
Since the sub-manifold defined by the image of
$\bs{\sigma}_{\eta}$ is by definition transversal to the action of
$\mr{PG}$, an infinitesimal gauge transformation of the gauge
fixing condition $\eta(\dot{\gamma}(t))=0$ will be always
non-trivial. In this way we can argue that the element $\Delta
[\gamma]$ is never zero. This fact ensures that the connection
$\eta$ induces a well-defined global gauge fixing, even when the
topology of the fiber bundle $\mr{A}\rightarrow \mr{A}/\mr{G}$ is
not trivial.

\vspace{0.5 cm}

By inserting~\eqref{eq:10} in~\eqref{kdu} we obtain
\begin{equation*}
  \langle A_0 \; \vert \; \pi^{-1}[A_1] \rangle = \int_{T^*\mathcal{P}(A_0,
    \, \pi^{-1}[A_1])}
  \, \da \mr{D}\pi \Delta [ \gamma ]  \int_{\mr{PG}}\mr{D} g'
  \,  \bs{\delta} (\eta
  (\dot{\gamma  g'}))) \exp\{i S\} \; .
  \end{equation*}

If we now perform in the usual manner a gauge transformation
taking $\gamma g'$ to $\gamma$ we obtain
\begin{equation*}
   \langle A_0 \; \vert \;  \pi^{-1}[A_1] \rangle =
   \int_{\mr{PG}}\mr{D} g' \int_{T^*\mathcal{P}(A_0, \,
     \pi^{-1}[A_1])}\mr{DA}\mr{D}\pi \Delta [ \gamma ]
  \,  \bs{\delta}  (\eta
    (\dot{\gamma}))) \exp\{i S\} \;,
  \end{equation*}
where we have used that $\mr{DA}\mr{D}\pi$, $\Delta^{-1} [ \gamma
]$ and $S$ are gauge invariant. In this way we have isolated the
infinite volume of the path group $\mr{PG}$.

\vspace{0.5 cm}

We will now follow the common procedure for finding the new terms
in the action coming from the Dirac delta $\bs{\delta} (\eta
(\dot{\gamma}))$ and the element $\Delta [\gamma]$.

In order to find an integral representation of the gauge
condition's Dirac delta $\bs{\delta}(\eta (\dot{\gamma}))$ we will
start by finding the integral representation of the Dirac delta
function $\delta_{\mathcal{L}ie(\mr{G})}$ on
$\mathcal{L}ie(\mr{G})$ used in (\ref{eq:9}). If $\xi\in
\mathcal{L}ie(\mr{G})$, the Dirac delta
$\delta_{\mathcal{L}ie(\mr{G})}(\xi)$ defined as
\begin{eqnarray*}
\delta_{\mathcal{L}ie(\mr{G})}(\xi) = \lim_{M} \prod^{M}_{j = 1}
\delta_{\liea} (\xi(x_{j})),
\end{eqnarray*}
can be expressed in terms of the integral representations of the
Dirac delta $\delta_{\liea} (\xi(x_{j}))$ on $\liea$ as
\begin{eqnarray*}
\delta_{\mathcal{L}ie(\mr{G})}(\xi) &=& \lim_{M} \prod^{M}_{j = 1}
\int d\lambda(x_{j}) e ^{i \sum ^{M}_{j = 1} \langle
\lambda(x_{j}), \xi(x_{j}) \rangle_{\liea}},
\\ &=& \int \widetilde{\mr{D}}\lambda e ^{i \int dx \langle \lambda(x),
\xi(x) \rangle_{\liea}},
\\ &=& \int \widetilde{\mr{D}}\lambda e ^{i \langle \lambda,
\xi \rangle_{\mathcal{L}ie(\mr{G})}},
\end{eqnarray*}
where $\lambda$ is a section of $\ad = P \times_G \liea$ and
$\widetilde{\mr{D}}\lambda = \lim_{M} \prod^{M}_{j = 1}
d\lambda(x_{j})$. The Dirac delta $\bs{\delta}(\eta
(\dot{\gamma}))$ of the gauge fixing condition can then be
expressed as
\begin{eqnarray*}
\bs{\delta} (\eta
    (\dot{\gamma})) &=& \lim_{N} \prod^{N}_{k = 1}
\delta_{\mathcal{L}ie(\mr{G})}(\eta (\dot{\gamma} (t_{k})))
\\ \nonumber &=& \lim_{N} \prod^{N}_{k = 1} \int
\widetilde{\mr{D}}\lambda_{k} e^{i\sum ^{N}_{k = 1} \langle
\lambda_{k}, \eta (\dot{\gamma} (t_{k})) \rangle_{\lieG}}
\\ \nonumber &=& \int \mr{D}\lambda e^{i \int_{\gamma} \langle
\lambda, \eta (\dot{\gamma} (t)) \rangle_{\lieG}dt}
\\ \nonumber &=& \int \mr{D}\lambda e^{i \int_{\gamma}  \int_M
\langle \lambda, \eta (\dot{\gamma} (t)) \rangle_\liea dx dt},
\end{eqnarray*}
where $\lambda$ is a time-evolving section of $\ad = P \times_G
\liea $. The final measure $\mr{D}\lambda$ is then
\begin{eqnarray*}
\mr{D}\lambda= \lim_{N} \prod^{N}_{k = 1}
\widetilde{\mr{D}}\lambda_{k} = \lim_{N, M} \prod^{N}_{k = 1}
\prod^{M}_{j = 1} d\lambda_{k}(x_{j}).
\end{eqnarray*}

Let's now compute explicitly the element $\Delta [\gamma]$. Let
$\mathbf{X}$ be an element of the Lie algebra
$\mathcal{L}ie(\mr{PG})$ identified with the tangent space of
$\mr{PG}$ at the identity element. Given a path $\gamma (t)\in
\mathcal{P}(A_0, \, \pi^{-1}[A_1])$, one must calculate the
variation of $\eta\left(\dot{\gamma}\right)$ under an
infinitesimal gauge transformation defined by $\mathbf{X} \in
\mathcal{L}ie(\mr{PG})$. Let $u\rightarrow k_{u}$ be the
uniparametric subgroup of $\mr{PG}$ generated by $\mathbf{X}$ by
means of the exponential map $\exp: \mathcal{L}ie(\mr{PG})
\rightarrow \mr{PG}$. We have then $\mathbf{X} = \tfrac{d}{du} k_u
\vert_{u=0} \in \mathcal{L}ie(\mr{PG})$.
\begin{rem}\label{sec:gener-gauge-fixing}
  The gauge fixing condition has a natural interpretation in terms of
  the geometry of $\mathcal{P} (A_0, \pi^{-1} [A_1])$. The tangent
  space $T_{\gamma} \mathcal{P} (A_0, \pi^{-1} [A_1])$ to $\mathcal{P}
  (A_0, \pi^{-1} [A_1])$ at a path $\gamma$ can be identified with the
  sections of the pullback $\gamma^* (TA)$. The connection $\eta$
  induces a map from $T_{\gamma} \mathcal{P} (A_0, \pi^{-1} [A_1])$ to
  the Lie algebra of $\mr{PG}$. The tangent field $\dot{\gamma}$
  represents a marked point in $T_{\gamma} \mathcal{P} (A_0, \pi^{-1}
  [A_1])$. The gauge fixing condition means that we will only consider
  paths such that the connection $\eta$ vanishes on the marked point
  $\dot{\gamma}$.
\end{rem}

Using the description of the tangent spaces to path spaces given
in the previous remark we can identify $\mathbf{X}$ with a map $t
\rightarrow X_t$ with $0 \leq t \leq 1$ and $X_t \in \lieG$. In
order to find an expression for the vectors $X_t$ one must take
into account that $k_u$ describes a family of paths in the gauge
group $\mr{G}$ parameterized by $t$. This means that
$k_u=g_{u}(t)\subset\mr{G}$ for $u$ fixed. To emphasize this, let
us write $k_u (t)$. For a given $t$, the vector $X_t \in \lieG$ is
then given by $X_t = \tfrac{d}{du} k_u (t) \big|_{u=0}$. If
$\gamma \in \mathcal{P}(A_0, \, \pi^{-1}[A_1])$ we must compute
\begin{equation}
  \label{eq:ku2}
  \frac{d}{du} \left(\eta_{\gamma (t) k_u(t)} \left( \frac{d}{dt} R_{k_u(t)}
    \gamma (t) \right) \right) \Big|_{u=0}.
\end{equation}

At a fixed time $t_0$ the time derivative in (\ref{eq:ku2}) is
equal to
\begin{eqnarray}
  \label{eq:122}
  \frac{d}{dt} R_{k_u(t)}
  \gamma (t) \Big|_{t=t_0} &=& \frac{d}{dt} R_{k_u(t_0)
    k_{u}(t_0)^{-1} k_{u}(t)} \gamma(t)
  \big\vert_{t=t_0}
  \\  \nonumber &=& \left(\frac{d}{dt} R_{k_{u}(t_0)} \gamma(t) \big\vert_{t=t_0} \right)
  \\ \nonumber &+& \frac{d}{dt} R_{k_{u}(t_0)^{-1} k_{u}(t)}
  \left(R_{k_u(t_0)}\gamma(t_0)\right)  \big\vert_{t=t_0}  \\
  \nonumber &=& dR_{k_{u}(t_0)} (\dot{\gamma} (t_0)) + \iota_{(\gamma(t_0) k_{u}(t_0))}
  (X_{u}),
\end{eqnarray}
where $X_{u}=k_{u}(t_0)^{-1} \frac{d}{dt} k_{u}(t) \Big|_{t=t_0}$.
The first term in the last equation is the differential of the
action of $k_u(t)$. We have used that the differential of a
function $f$ is defined as
$df(X)=\frac{df(\gamma(t))}{dt}\Big|_{t=t_0}$ with
$X=\frac{d\gamma(t)}{dt}\Big|_{t=t_0}$. The second term is the
homomorphism from the Lie algebra of $\mr{G}$ to the vertical
vector fields (see the appendix~\ref{appendix} for detailed
definitions).\footnote{If we use the usual form for the gauge
transformation of connections~\eqref{eq:ap20} the time derivative
in~\eqref{eq:ku2} can be expressed as
\begin{displaymath}
  \text{ad} (k_u^{-1}(t_0)) \dot{\gamma} (t_0)+ \frac{d}{dt} \left(\text{ad}
    \left(k_u^{-1}(t) \right) \gamma (t_0)  + k_u^{-1}(t) dk_u(t)
    \right)\Big|_{t=t_0}.
\end{displaymath}

The first term is equal to the differential of the action
$dR_{k_{u}(t_0)} (\dot{\gamma} (t_0))$ (see (\ref{eq:ap60})). The
second term is equal to the infinitesimal symmetry
$\iota_{(\gamma(t_0) k_{u}(t_0))} (X_{u})$ (see (\ref{eq:ap50})).}

Let's apply the connection form $\eta $ to each term in
equation~\eqref{eq:122}. Firstly, we have
\begin{equation}
  \label{eq:13}
  \eta_{\gamma(t_0) k_{u}(t_0)} (dR_{k_{u}(t_0)} (\dot{\gamma} (t_0))) = \text{Ad}
  (k_{u}^{-1}(t_0)) \eta_{\gamma(t_0)} (\dot{\gamma} (t_0)).
\end{equation}

The equality~\eqref{eq:13} follows from one of the connection's
defining properties (see~\eqref{eq:eta2} in the
appendix~\ref{appendix}). The second term is
\begin{equation}
  \label{eq:etavert}
  \eta_{\gamma(t_0) k_{u}(t_0)} \left( \iota_{(\gamma(t_0)
      k_{u}(t_0))} (X_{u}) \right) = X_u
\end{equation}
where we have used equation~\eqref{eq:eta-properties}. The
infinitesimal variation defined by $\mathbf{X}\in
\mathcal{L}ie(\mr{PG})$ is given by the sum of
\begin{eqnarray*}
\frac{d}{du}\left(\text{Ad} (k_{u}(t_0)^{-1}) \eta_{\gamma(t_0)}
(\dot{\gamma} (t_0))  \right)\Big|_{u=0} &=& \text{Ad} (-X_{t_0})
\eta_{\gamma(t_0)} (\dot{\gamma} (t_0))
\\ &=& \left[-X_{t_0},\eta_{\gamma(t_0)} (\dot{\gamma} (t_0)) \right]
\end{eqnarray*}
and $\tfrac{d}{du} X_{u}|_{u=0}$. Let's now calculate this last
term:
\begin{eqnarray*}
\frac{d}{du} X_{u}|_{u=0} &=& \frac{d}{du} \left(k_{u}(t_0)^{-1}
\frac{d}{dt} k_{u}(t)  \right)\Big|_{u=0, t=t_0}
\\ &=& \left(-k_{u}(t_0)^{-2}
  \frac{dk_{u}(t_0)}{du}\frac{dk_{u}(t)}{dt}+k_{u}(t_0)^{-1}\frac{d}{dt}\frac{dk_{u}(t)}{du}\right) \Big|_{u=0, t=t_0}
\\ &=& -k_{0}(t_0)^{-2} \frac{dk_{u}(t_0)}{du}\Big|_{u=0}
\frac{dk_{0}(t)}{dt}\Big|_{t=t_0} \\ & &
+k_{0}(t_0)^{-1}\frac{d}{dt}\left[\frac{dk_{u}(t)}{du}\Big|_{u=0}\right]_{t=t_0}
\\ &=&  \dot{X}_{t}(t_0),
\end{eqnarray*}
where in the last step we used that $k_{0}(t)=id_{\mr{G}}$
$\forall t$.

The infinitesimal variation of the gauge fixing condition is then

\begin{eqnarray*}
\frac{d}{du} \left(\eta_{\gamma (t) k_u(t)} \left( \frac{d}{dt}
R_{k_u(t)} \gamma (t) \right) \right) \Big|_{u=0}=
\left[-X_{t},\eta_{\gamma(t)} (\dot{\gamma} (t)) \right]+
\dot{X}_{t}.
\end{eqnarray*}

This expression defines a linear endomorphism $M_{\gamma}$ in
$\mathcal{L}ie(\mr{PG})$ for each path $\gamma(t)$. Equivalently,
it defines a linear endomorphism $M_{\gamma}(t)$ in $\lieG$ for
each $t$.

In order to find the exponential representation of the element
$\Delta\left[\gamma\right]$, we will introduce a Grassmann algebra
generated by the anticommuting variables $c$ and $\bar{c}$. By
following the common procedure, we can express the element
$\Delta\left[\gamma\right]$ as
\begin{eqnarray*}
\Delta[\gamma] &=& \int \mr{D}\bar{c} \mr{D}c e^{\int \langle
\bar{c}_{t},M_{\gamma}(t) c_{t} \rangle_{\liea}d^{3}xdt},
\end{eqnarray*}
where $$\mr{D}c= \lim_{N} \prod^{N}_{k = 1}
\widetilde{\mr{D}}c_{t_{k}} = \lim_{N, M} \prod^{N}_{k = 1}
\prod^{M}_{j = 1} dc_{t_{k}}(x_{j}),$$ and the same for
$\mr{D}\bar{c}$ (see Ref.\cite{witten99:_dynam_quant_field_theo}
for a precise definition of $c$ and $\bar{c}$).

By gathering all the pieces together, the path integral takes the
form
\begin{equation} \label{gfpi}
   \langle A_0 \vert \pi^{-1}[A_1] \rangle = \int_{\mr{PG}}\mr{D} g \int \mr{DA}\mr{D}\pi \mr{D}\lambda \mr{D}c \mr{D}\bar{c}
  \,  \exp\{i S_{gf} \},
  \end{equation}
where $S_{gf}$ is the gauge fixed action
\begin{eqnarray*}
S_{gf}=\int d^{4}x\left(\dot{A}_{k}\pi^{k} - \mathcal{H}_{0}-A_{0}
\phi + \langle \lambda, \eta (\dot{\gamma}) \rangle_{\liea} - i
\langle \bar{c}, M_{\gamma}c \rangle_{\liea} \right).
\end{eqnarray*}

By explicitly introducing the indices of the Lie algebra $\liea$,
the endomorphisms $M_{\gamma}(t)(x)$ can be expressed as
\begin{equation*}
 \label{eq:28}
M_{\gamma}(t)(x)_{a}^{c}=-\eta_{\gamma(t)} (\dot{\gamma} (t))^{b}
f_{ab}^{c}+\delta_{a}^{c}\partial_{0},
\end{equation*}
where $f_{ab}^{c}$ are the structure constants of $\liea$. The
gauge fixed action takes then the form
\begin{eqnarray} \label{jcp}
\\ \nonumber S_{gf}=\int d^{4}x\left(\dot{A}^{a}_{k}\pi^{k}_{a} -
\mathcal{H}_{0}-A_{0}^{a} \phi_{a} + \lambda_{a}
\eta(\dot{\gamma})^{a} + i
\bar{c}^{a}\eta(\dot{\gamma})^{b}f_{ab}^{c}c_{c}
-i\bar{c}^{a}\dot{c}_{a} \right).
\end{eqnarray}

The last term can be recast as $+i\dot{c}_{a} \bar{c}^{a}$.
Therefore, this term can be interpreted as the kinetic term
corresponding to the new pair of canonical variables $\left(c,
i\bar{c}\right)$.

\section{Conclusions}
\label{sec:5}

The principal aim of this work was to study the quantization of
Yang-Mills theory by using an extended connection $\AA$ defined in
a properly chosen principal bundle. This connection unifies the
three fundamental geometric objects of Yang-Mills theory, namely
the gauge field, the gauge fixing and the ghost field. This
unification is an extension of the known fact that the gauge and
ghost fields can be assembled together as $\omega = A + c$
\cite{Baulieu-Singer1988}.

The first step in the unification process was to generalize the
gauge fixing procedure by replacing the usual gauge fixing section
$\sigma$ with a gauge fixing connection $\eta$ in the
$\mr{G}$-principal bundle $\mr{A} \rightarrow \mr{A}/\mr{G}$. We
have then shown that the connection $\eta$ also encodes the ghost
field of the BRST complex. In fact, the ghost field can either be
considered the canonical vertical part of $\eta$ in a local
trivialization or the universal connection in the gauge group's
Weil algebra. The unification process continues by demonstrating
that the universal family of gauge fields $\mathbf{A}^{U}$ and the
gauge fixing connection $\eta$ can be unified in the single
extended connection $\AA=\mathbf{A}^{U} + \eta$ on the
$G$-principal bundle $\mr{A} \times P \rightarrow \mr{A} \times
M$. In this way, we have shown that the extended connection $\AA$
encodes the gauge field, the gauge fixing and the ghost field. A
significant result is that it is possible to derive the BRST
transformations of the relevant fields without imposing the usual
horizontality or flatness conditions on the extended curvature
$\FF=\phi+\psi+\mathbf{F}^{U}$ (\cite{Baulieu-Bellon},
\cite{Baulieu-Mieg1982}, \cite{Mieg}). In other words, it is not
necessary to assume that $\phi=\psi=0$. Moreover, the proposed
formalism allows us to show that the standard BRST transformation
of the gauge field $A$ is only valid in a local trivialization of
the fiber bundle $\mr{A} \rightarrow \mr{A}/\mr{G}$. In fact,
equation \eqref{brstA} can be considered the corresponding global
generalization.

We then applied this geometric formalism to the path integral
quantization of Yang-Mills theory. Rather than selecting a fixed
representant for each $[A] \in \mr{A}/\mr{G}$ by means of a
section $\sigma$, the gauge fixing connection $\eta$ allows us to
parallel transport any initial condition $A_{0}\in \mr{A}$
belonging to the orbit $[A_{0}]$. A significant advantage of this
procedure is that one can always define a section
$\bs{\sigma}_{\eta}$ of the projection $\mathcal{P} (A_0,\pi^{-1}
[A_1]) \rightarrow \mathcal{P} ([A_0], \, [A_1])$ in the space of
paths, even when it is not possible to define a \emph{global}
section $\sigma: \mr{A}/\mr{G} \rightarrow \mr{A}$ in the space of
fields. Since the path integral is not an integral in the
\emph{space of fields} $\mr{A}$, but rather an integral in the
\emph{space of paths} in $\mr{A}$, such a section
$\bs{\sigma}_{\eta}$ suffices for eliminating the infinite volume
of the group of paths $\mr{PG}$. Hence, this generalized gauge
fixing procedure is globally well-defined even when the topology
of the fiber bundle $\mr{A} \rightarrow \mr{A}/\mr{G}$ is not
trivial (Gribov's obstruction).

We then used the standard Faddeev-Popov method in order to
introduce the generalized gauge fixing defined by $\eta$ at the
level of the path integral. The corresponding gauge fixed extended
action $S_{gf}$ was thereby obtained. We have thus shown that the
Faddeev-Popov method can be used even when there is a Gribov's
obstruction.

\vspace {0.5 cm}

\textbf{Acknowledgement}

\vspace {0.25 cm}

We wish to thank Marc Henneaux for his helpful comments and the
University of Buenos Aires for its financial support (Projects No.
X103 and X193).

\newpage
\appendix

\section{The geometry of $\eta$}\label{appendix}

In this appendix we will review some geometric properties of the
connection $\eta$. The gauge group $\mr{G}$ consists of
diffeomeorphisms $\varphi:P \rightarrow P$ such that $ \pi \varphi
= \pi$ and $\varphi(pg) = \varphi(p) g$ (with $g\in G$). Each
element $\varphi \in \mr{G}$ can be associated to a map
$\mathbf{g}: P \rightarrow G$ by $\varphi(p) = p \mathbf{g}(p)$.
This map $\mathbf{g}$ satisfies $\mathbf{g}(ph) = h^{-1}
\mathbf{g} (p) h = \text{ad} (h^{-1}) \mathbf{g}(p)$. This
description of the elements of $\mr{G}$ also allows one to
describe the elements of the Lie algebra $\lieG$. These elements
consist of maps $\boldsymbol{\mathfrak{g}}: P \rightarrow
\mathfrak{g}$ such that $\boldsymbol{\mathfrak{g}} (ph) =
\text{Ad} (h^{-1}) \boldsymbol{\mathfrak{g}} (p)$.

The gauge group $\mr{G}$ acts on the right on $\mr{A}$ via the
pullback of connections. If $\varphi: P \rightarrow P $ is an
element of $\mr{G}$ and $A \in \mr{A}$, then the action is given
by:
\begin{equation}
  \label{eq:ap10}
  R_{\varphi} A = \varphi^* A.
\end{equation}

This action is commonly described in terms of the function
$\mathbf{g}$ by the transformation formula
\begin{equation}
  \label{eq:ap20}
  R_{\varphi} A = \mathbf{g}^{-1} A \mathbf{g} + \mathbf{g}^{-1} d\mathbf{g}.
\end{equation}

The first term in the right hand side of equation~\eqref{eq:ap20}
denotes the composition
\begin{equation}
  \label{eq:ap30}
  T_pP \xrightarrow{A_p} \liea \xrightarrow{ \text{ad}_{g^{-1} (p)}}
  \liea.
\end{equation}

The second part $\mathbf{g}^{-1} d\mathbf{g} $ is the pullback of
the Maurer-Cartan form $\omega $ defined by the map $\mathbf{g}: P
\rightarrow G$.

The action~\eqref{eq:ap10} induces two constructions of interest.
The first one is the \emph{infinitesimal action.} This is a
morphism of Lie algebras $\iota : \lieG \rightarrow \Gamma(TP)$,
where $\Gamma (TP)$ is the Lie algebra of the vector fields on
$P$. We will identify elements of a Lie algebra with tangent
vectors at the identity. If $\zeta \in \lieG$ and $\varphi_s$ is a
curve such that
$$
\frac{d}{ds} \varphi_s \Big|_{s=0} = \zeta,
$$
and $A \in \mr{A}$, then we define
\begin{equation}
  \label{eq:ap40}
  \iota_A \zeta = \frac{d}{ds} R_{\varphi_s} A \Big |_{s=0}.
\end{equation}

In terms of the explicit description of equation~\eqref{eq:ap20}
we obtain
\begin{equation}
  \label{eq:ap50}
  \iota_A \zeta = \text{Ad} (\zeta) A + \frac{d}{ds} \mathbf{g}^{-1}
  d\mathbf{g} \Big |_{s=0}.
\end{equation}

The second action is the differential of the action of $\mr{G}$.
Since $\mr{A}$ is an affine space, there is a natural
identification $T \mr{A} = \mr{A} \oplus \mr{A}$. An element $V
\in T_A \mr{A}$ can then be identified with a connection. Let
$A_s$ be a curve in $\mr{A}$ such that $\tfrac{d}{ds} A_s
\vert_{s=0} = V$. Then we define
\begin{equation}
  \label{eq:ap60}
  dR_{\varphi} V = \frac{d}{ds} \text{ad} (\mathbf{g}) A_s \Big|_{s=0}.
\end{equation}
\begin{rem}
  The infinitesimal action and the differential of the action are
  \emph{geometric intrinsic constructions.} They only depend on the
  vectors $\zeta$ and $V$ and not on the particular curves used to
  compute them.
\end{rem}
The connection $\eta$ has the following properties
\begin{eqnarray}
\label{eq:eta-properties} \eta(\iota (\zeta)) &=& \zeta,
\\ \label{eq:eta2}
    \eta(dR_{\varphi} V) &=& \text{ad} (\mathbf{g}^{-1}) \eta(V).
\end{eqnarray}

\section{The local form of $\eta$}
\label{sec:local-form-eta}

The local form of a connection in a principal bundle is commonly
described in terms of \emph{Christoffel symbols}. This description
has been extended to infinite dimensions in Ref.\cite[Chapter
VIII]{kriegl97}. We will give a brief description of the
constructions involved. Let $U_{\alpha}$ be a trivializing
covering of $\mr{A}/\mr{G}$ with trivializing functions
$\psi_{\alpha}: U_{\alpha} \times \mr{G} \rightarrow \mr{A}
\big|_{U_{\alpha}}$. Then the pullback of $\eta$ defines a
$\lieG$-valued 1-form on $U_{\alpha} \times \mr{G}$. Let $v \in
T_{x} U_{\alpha}$ and $w \in T_\mathbf{g} \mr{G}$ be two tangent
vectors. Then
\begin{equation}
  \label{eq:ap100}
  \psi_{\alpha}^* \eta (v, w) =: - \Gamma^{\alpha} (v, \mathbf{g}) + w
\end{equation}
for a certain 1-form $\Gamma^{\alpha}$ with values on the vector
fields on $\mr{G}$. The 1-form $\Gamma^{\alpha}$ is called the
\emph{local Christoffel form} of $\eta$ in the trivialization
$\left(U_{\alpha}, \psi_{\alpha}\right)$. If $\gamma$ is a path in
$U_{\alpha}$, then any lift $\tilde{\gamma}$ of $\gamma$ to
$U_{\alpha} \times \mr{G}$ has the form $\tilde{\gamma} (t) =
(\gamma(t), \tau(t))$ for a path $\tau$ in $\mr{G}$. Then the
gauge fixing condition is given in local coordinates by the
equation
\begin{equation}
  \label{eq:final}
  \frac{d}{dt} \tau(t) - \Gamma^{\alpha} (\frac{d}{dt} \tau(t), \gamma
  (t)) = 0.
\end{equation}

\bibliographystyle{plain}

\begin{thebibliography}{99}
\bibitem{atiyah84:_dirac}
M.~F.~Atiyah and I.~M.~Singer, \emph{{D}irac operators coupled to
vector potentials}, \emph{Proc. National Acad. Sci.} \textbf{81}
(1984) 2597.

\bibitem{Baulieu-Bellon}
L.~Baulieu and M.~Bellon, \emph{p-forms and Supergravity: Gauge
symmetries in curved space}, \emph{Nucl. Phys.} \textbf{B}
\textbf{266} (1986) 75.

\bibitem{Baulieu-Singer1988}
L.~Baulieu and I.~M. Singer, \emph{Topological {Y}ang-{M}ills
symmetry}, \emph{Nucl. Phys.} \textbf{5} (\emph{Proc. Suppl.})
(1988) 12.

\bibitem{Baulieu-Mieg1982}
L.~Baulieu and J.~Thierry-Mieg, \emph{The principle of BRST
symmetry: An alternative approach to Yang-Mills theories},
\emph{Nucl. Phys.} \textbf{B} \textbf{197} (1982) 477.

\bibitem{Blau91}
D.~Birmingham, M.~Blau, M.~Rakowski and G.~Thompson,
\emph{Topological Field Theory}, \emph{Phys. Rep.} \textbf{209}
(1991) 129.

\bibitem{bonora83}
L.~Bonora and P.~Cotta-Ramusino, \emph{Some remarks on {BRS}
transformations, anomalies and the cohomology of the lie algebra
of the group of gauge transformations}, \emph{Commun. Math. Phys.}
\textbf{87} (1983) 589.

\bibitem{Choquet-Bruhat}
Y.~Choquet-Bruhat and C.~DeWitt-Morette, \emph{Analysis, Manifolds
and Physics. Part II: 92 Applications}, Elsevier Science
Publishers B.V., New York (1989).

\bibitem{cordes}
S.~Cordes, G.~Moore and S.~Ramgoolam, \emph{Lectures on 2D
Yang-Mills Theory, Equivariant Cohomology and Topological Field
Theories}, \emph{Nucl. Phys.} \textbf{41} (\emph{Proc. Suppl.})
(1995) 184.

\bibitem{donaldson90}
S.~K.~Donaldson and P.~B.~Kronheimer, \emph{The geometry of
four-manifolds}, Oxford University Press (1990).

\bibitem{dubois}
M.~Dubois-Violette, \emph{The Weil-B.R.S. algebra of a Lie algebra
and the anomalous terms in gauge theory}, \emph{J. Geom. Phys.},
\textbf{3} (1986) 525.

\bibitem{faddeev91:_gauge}
L.~Faddeev and A.~Slavnov, \emph{Gauge fields: An introduction to
quantum theory}, second ed., Frontiers in Physics, Perseus Books
(1991).

\bibitem{Gribov:1977wm}
V.~Gribov, \emph{Quantization of non-Abelian gauge theories},
Nucl. Phys. \textbf{B} \textbf{139} (1978) 1.

\bibitem{guillemin99}
V.~Guillemin, S.~Sternberg, and V.~W.~Guillemin,
\emph{Supersymmetry and equivariant de {R}ham theory},
Spinger-Verlag (1999).

\bibitem{henneaux85}
M.~Henneaux, \emph{Hamiltonian form of the path integral for
theories with a gauge freedom}, \emph{Phys. Rep.} (1985) 126.

\bibitem{henneaux94}
M.~Henneaux and C.~Teitelboim, \emph{Quantization of gauge
systems}, Princeton Univ. Press (1994).

\bibitem{kobayashi63:_found}
S.~Kobayashi and K.~Nomizu, \emph{Foundations of differential
geometry}, vol.~I, Wiley, New York (1963).

\bibitem{kriegl97}
A.~Kriegl and P.~Michor, \emph{A convenient setting for global
analysis}, Mathematical Surveys and Monographs, vol.~53, A.M.S
(1997).

\bibitem{Michor91}
P.~Michor, \emph{Gauge theory for fiber bundles}, Monographs and
Textbooks in Physical Sciences, Lecture Notes \textbf{19},
Bibliopolis, Napoli (1991).

\bibitem{Narasimhan79}
M.~S.~Narasimhan and T.~R.~Ramadas, \emph{Geometry of $SU(2)$
Gauge Fields}, \emph{Commun. Math. Phys.} \textbf{67} (1979) 121.

\bibitem{singer78}
I.~Singer, \emph{Some remarks on the Gribov Ambiguity},
\emph{Commun. Math. Phys.} \textbf{60} (1978) 7.

\bibitem{szabo00}
R.~Szabo, \emph{Equivariant cohomology and localization of path
integrals}, Springer-Verlag (2000).

\bibitem{Mieg}
J.~Thierry-Mieg, \emph{Geometrical reinterpretation of
Faddeev-Popov ghost particles and BRS transformations}, \emph{J.
Math. Phys.} \textbf{21} (1980) 2834.

\bibitem{Witten:1988ze}
E.~Witten, \emph{Topological quantum field theory}, \emph{Commun.
Math. Phys.} \textbf{117} (1988) 353.

\bibitem{witten99:_dynam_quant_field_theo}
E.~Witten, \emph{Dynamics of Quantum Field Theory} in Quantum
Fields and Strings: A course for mathematicians. Vol(2) 1119-1424
A.M.S (1999).
\end{thebibliography}
\providecommand{\bysame}{\leavevmode\hbox
to3em{\hrulefill}\thinspace}

\end{document}